\documentclass[12pt,draftcls,onecolumn]{IEEEtran}

\makeatletter
\def\ps@headings{%
\def\@oddhead{\mbox{}\scriptsize\rightmark \hfil \thepage}%
\def\@evenhead{\scriptsize\thepage \hfil \leftmark\mbox{}}%
\def\@oddfoot{}%
\def\@evenfoot{}}
\makeatother

\usepackage{amstext,amsmath,amssymb}
\usepackage{amsthm}
\usepackage{times}
\usepackage{graphicx}
\usepackage{algorithmic}
\usepackage{algorithm}
\usepackage{url}
\usepackage{graphicx}
\usepackage{color}
\usepackage{verbatim}
\usepackage{cite}

\newtheorem{theorem}{Theorem}
\newtheorem{lemma}[theorem]{Lemma}
\newtheorem{corollary}[theorem]{Corollary}
\newtheorem{definition}{Definition}

\newcounter{assump}

\newtheorem{assumption}[assump]{\bf AS}

\newcommand{\bdm}{
    \begin{displaymath}}

\newcommand{\edm}{
    \end{displaymath}}

\newcommand{\be}{
    \begin{equation}}

\newcommand{\ee}{
    \end{equation}}

\newcommand{\bea}{
    \begin{eqnarray}}

\newcommand{\eea}{
    \end{eqnarray}}

\newcommand{\beit}{\begin{itemize}}
\newcommand{\eeit}{\end{itemize}}

\newcommand{\eat}[1]{ }

\renewcommand{\algorithmiccomment}[1]{// {\em #1}}

\newcommand{\mainspacing}{1.38}

\renewcommand{\baselinestretch}{\mainspacing}

\begin{document}
%
\title{Algorithms and Stability Analysis for
Content Distribution over Multiple Multicast Trees}


\author{Xiaoying~Zheng,
        Chunglae~Cho,
        and~Ye~Xia
\thanks{X. Zheng is with Shanghai Advanced Research Institute, Chinese Academy of Sciences and Shanghai Research Center for Wireless Communications, China. E-mail: zhengxy@sari.ac.cn.}%
\thanks{C. Cho and Y. Xia are with the Department of Computer and Information Science and Engineering, University of Florida, Gainesville, FL 32611. E-mail: \{ccho, yx1\}@cise.ufl.edu.}
}

\maketitle

\begin{abstract}
The paper investigates theoretical issues in applying the
universal swarming technique to efficient content distribution.
In a swarming session, a file is distributed to all the receivers by
having all the nodes in the session exchange file chunks. By
universal swarming, not only all the nodes in the session, but
also some nodes outside the session may participate in the chunk
exchange to improve the distribution performance. We present a
universal swarming model where the chunks are distributed along
different Steiner trees rooted at the source and covering all the
receivers. We assume chunks arrive dynamically at the sources and
focus on finding stable universal swarming algorithms. To achieve
the throughput region, universal swarming usually involves a
tree-selection subproblem of finding a min-cost Steiner tree,
which is NP-hard. We propose a universal swarming scheme that
employs an approximate tree-selection algorithm. We show that it
achieves network stability for a reduced throughput region, where
the reduction ratio is no more than the approximation ratio of the
tree-selection algorithm. We propose a second universal swarming
scheme that employs a randomized tree-selection algorithm. It
achieves the throughput region, but with a weaker stability
result. The proposed schemes and their variants are expected to be useful for infrastructure-based content distribution networks with massive content and relatively stable network environment.
\end{abstract}



\section{Introduction}
\label{sec:introduction}

The Internet is being used to transfer content on a more and more
massive scale. A recent innovation for efficient content
distribution is a technique known as {\em swarming}. In a swarming
session, the file to be distributed is broken into many chunks at
the original source, which are then spread out across the
receivers. Subsequently, the receivers exchange the chunks with
each other to speed up the distribution process. Many different
ways of swarming have been proposed, such as FastReplica
\cite{LeeV07}, Bullet \cite{Bullet'}, Chunkcast \cite{Chunkcast},
BitTorrent \cite{BitTorrent}, and CoBlitz \cite{CoBlitz}.


The swarming technique was originally introduced by the end-user communities for peer-to-peer (P2P) file sharing. The subject of this paper is how to apply swarming to infrastructure-based content distribution, where files are to be distributed among content servers in  a content delivery network. The content servers are usually connected with leased high-speed links and, as a result, the bandwidth bottleneck may no longer be at the access links. It has been shown that content delivery traffic has made capacity shortage in the backbone networks a genuine possibility \cite{CFE06}. The size of such a content distribution network is usually small, consisting of up to hundreds of network nodes and up to thousands of servers. Unlike the dynamic end-user file-sharing situations, infrastructure networks and content servers are usually centrally managed, generally well-behaved and relatively static (however, the traffic can still be dynamic).
In this setting, it is beneficial to view swarming as distribution over multiple multicast trees, each spanning the content servers. This view allows us to pose the question of how to optimally distribute the content (see \cite{ZCX08}). Furthermore, it is often easier to first develop sophisticated algorithms under this tree-based view, and then, adapt them to practical situations where the tree-based view is only partially adequate. Hence, in this paper, swarming is synonymous to distribution over multiple multicast trees.

\eat{

\textcolor{red}{
Compared with the dynamic end-user
file-sharing situation, the infrastructure networks and the
content servers are usually centrally managed,
generally well-behaved, relatively static
(however, the traffic can
still be dynamic).
They are connected with leased high-speed overlay
links, which are no longer access link bandwidth
constrained.
Content delivering traffic has made capacity shortage in
the backbone networks a genuine possibility, which will
become more serious with the increasing deployment
of fiber-based access.
\footnote{At the present, telecom
companies are aggressively deploying fiber-to-the home (FTTH) or
its variants. Verizon is building a nationwide FTTH network in the
US, to be completed by 2010. Japan had 5.6 million FTTH
subscribers in 2006 and will have 30 million by 2010
\cite{Wiel06}. The fiber speed is currently at up to 100 Mbps,
heading to 1 Gbps by 2020 and will reach 10 Gbps later
\cite{Geo05}.}
As an example, with its early adoption of FTTH, by
2005, Japan already saw that the fiber users were responsible
for 86\% of the inbound traffic; and the traffic was rapidly
increasing, by 45\% that year \cite{CFE06}.}
\eat{As an example, with its early adoption of FTTH, by
2005, Japan already saw 62\% of its backbone network traffic being
from residential users to users, which was consumed by content
downloading or P2P file sharing; the fiber users were responsible
for 86\% of the inbound traffic; and the traffic was rapidly
increasing, by 45\% that year \cite{CFE06}.}

}

This paper concerns a class of improved swarming techniques, known
as {\em universal swarming}. We associate with each file to be
distributed a {\em session}, which consists of the source of a
file and the receivers who are interested in downloading the file.
In traditional swarming, chunk exchange is restricted to the nodes
of the session. However, in \emph{universal swarming}, multiple
sessions are combined into a single ``super session'' on a shared
overlay network. Universal swarming takes advantages of the
heterogenous resource capacities of different sessions, such as
the source upload bandwidth, receiver download bandwidth, or
aggregate upload bandwidth, and allows the sessions to share each
other's resources. The result is that the distribution efficiency
of the resource-poor sessions can improve greatly with negligible
impact on the resource-rich sessions (see \cite{ZCX09-1a}).

\begin{figure}[t]
\begin{center}
\includegraphics[width=2in]{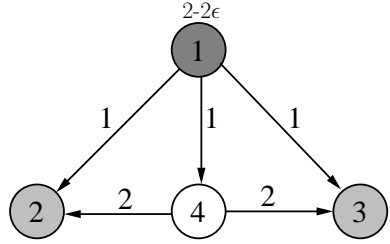}
\end{center}
\caption{Node 1 sends the file to nodes 2 and 3.}
\label{fig:three_nodes}
\end{figure}

\begin{figure}[t]
\begin{center}
\includegraphics[width=3in]{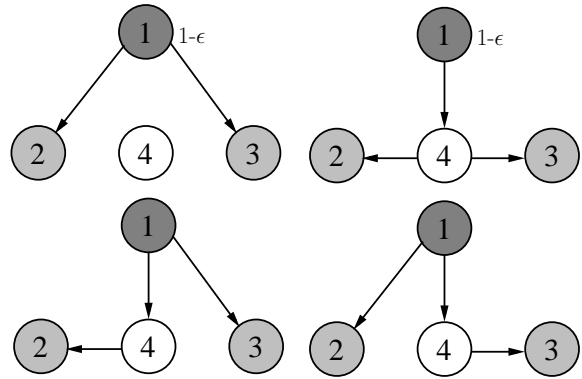}
\end{center}
\caption{All possible distribution trees for the example in Fig.
\ref{fig:three_nodes}.} \label{fig:dist_tree}
\end{figure}

In universal swarming, if we focus on a particular file, not only
the source and all the receivers participate in the chunk exchange
process, some other nodes who are not interested in the file may
also participate. We call the latter out-of-session nodes. To
illustrate the essence of universal swarming, as well as the main
issues, consider the toy example in Fig. \ref{fig:three_nodes}.
The numbers associated with the links are their capacities. Let us
consider a particular file for which the source is node 1 and the
receivers are nodes 2 and 3. Node 4 is out of the session. Let us
focus on a fixed chunk and consider how it can be distributed to
the receivers. With some thoughts, it can be seen that the chunk
propagates on a tree rooted at the source and covering both
receivers. All possible distribution trees are shown in Fig.
\ref{fig:dist_tree}. We notice that a distribution tree may or may
not include the out-of-session node, 4. Thus, a distribution tree
in general is a {\em Steiner} tree rooted at the source covering
all the receivers, where the out-of-session nodes (e.g., node 4)
are the Steiner nodes.

With this model of multi-tree multicast, one of the main questions
is how to assign the chunks to different distribution trees so as
to optimize certain performance objective, such as maximizing the
sum of the utility functions of the sessions, or minimizing the
distribution time of the slowest session. This is a {\em rate
allocation problem} on the multicast trees. One such question was
addressed in \cite{ZCX08} in the context of non-universal
swarming, where each session's multicast trees are spanning trees
instead of Steiner trees. For universal swarming, the question was
addressed in \cite{ZCX09-1a}.

This paper addresses the {\em stability problem}. The main
question is: Given a set of data rates from the sources (which are
possibly the solutions to the aforementioned rate allocation
problem), how do we get a universal swarming algorithm so that the
network queues will be stable?
For the example in Fig. \ref{fig:three_nodes}, a source rate of 2 is the largest distribution rate that can be supported by the network if everything is deterministic. To achieve stability under random arrivals, it usually requires that the data arrival rate is strictly less than 2 (for a justification, consider a single-queue system).
Hence, when the
file chunks arrive at (or generated by) the source node 1 at a
mean rate $2 - 2 \epsilon$, where $0 < \epsilon
< 1$, we can place chunks on the first and
the second tree in Fig. \ref{fig:dist_tree} at a mean rate $1 -
\epsilon$ each. For this example, the solution actually stabilizes
the network. But, this conclusion requires technical conditions
and is not generally true for more complicated situations.


In this paper, we develop a universal swarming scheme that employs
an approximation algorithm to the tree selection ({\em a.k.a.}
scheduling) problem, which achieves a rate region\footnote{Subsequently, when we say an algorithm achieves or
stabilizes a region, we mean the interior of the region.} equal to the
throughput region reduced by a constant factor $\gamma$, $\gamma
\geq 1$. We show that $\gamma$ is no more than the approximation
ratio of the tree scheduling algorithm. The scheme requires
network signaling and source traffic regulation. We propose a
second universal swarming scheme that utilizes a randomized tree
selection algorithm, which achieves the entire throughput region,
but with a weaker stability property.

The difference between our problem and the wireless scheduling problems is substantial.
Most previous papers that consider multi-hop traffic are either in the unicast setting or in a multicast setting with a few fixed multicast trees per multicast session.
We must consider multi-hop, multicast communications, and, to have the largest possible throughput region, we allow each multicast session to use {\em any} multicast trees for the session. The combination of these three features makes our problem both unique and very hard. One of the main challenges is that there is no obvious way to specify the packet forwarding behavior without per-tree queueing (i.e., having a separate queue for each multicast tree) whereas, on the other hand, per-tree queueing is impractical due to the exceedingly large number of trees. Our solution to resolve this difficulty, on the algorithm side, is to use the techniques of signaling and virtual queueing. The techniques allow us to have very small numbers of queues and at the same time show stability guarantee.

Having avoided per-tree queueing, however, the performance analysis is still difficult. In particular, there is no easy way to prove the stability of the real queues.
Our analytical approach is to first prove the stability of the virtual queues. The proofs in this step are relatively conventional, using the techniques of Lyapunov drift analysis. The second step is to make connections between the virtual queues and the real queues, and use the stability results for the virtual queues to prove that the real queues are also stable.

The tree selection subproblem is inherent to the problem formulated in the paper and it remains difficult. As a remedy, our algorithms can work with low-cost trees as opposed to the min-cost trees; finding low-cost trees can be much easier.
The stability results are applicable to classes of algorithms by allowing different tree selection sub-algorithms. Hence, the research for finding simpler, more practical algorithms can continue.

The rest of the paper is organized as follows. The models and the problem
description are given in Section \ref{sec:models_stochastic}.
The first universal swarming scheme and the analysis
are presented in Section \ref{sec:regulator}.
The second universal swarming scheme and the analysis
are presented in Section \ref{sec:without_regulator}.
In Section \ref{sec:related}, we discuss additional related work.
The conclusion is in Section \ref{sec:conclusion}.


\section{Problem Description}
\label{sec:models_stochastic}

We consider a time-slotted system where each time slot has a
duration of one time unit. Let the network be represented by a
directed graph $G = (V, E)$, where $V$ is the set of nodes and $E$
is the set of links. For each link $e \in E$, let $c_e$ denote its
capacity (e.g., in the number
of file chunks it can transmit per time slot),
where $c_e > 0$. We assume that each session, which distributes
a distinct file, has one source, and hence, there is a one-to-one
mapping between a session and a source.\footnote{The case of multiple sources makes the most sense when the sources each possess a copy of a common file, which will be distributed to a common group of receivers. We can extend the network graph by adding a virtual node and virtual edges. Each virtual edge connects the virtual node to one of the sources, and it is given an infinite capacity. In the expanded graph, the virtual node will be considered the source of the multicast session. In the dynamic case where chunks of a common file arrive at the sources following stochastic processes, it is difficult to parsimoniously specify the relationship among the file chunks to different sources. Without that information, we will identify each source as a separate session (which has the same set of receivers).}
Let $S$ denote the set of sources (sessions). For each $s \in S$,
let $V_s \subseteq V$ be the set of receivers associated with the
source (session) $s$.

For each source $s \in S$, suppose constant-sized data packets (i.e., file
chunks)\footnote{Since our algorithms can be deployed more easily at the application level, in this paper, we use the term {\em data packet} (or {\em real packet}) to mean an application-level data unit and we regard it as being synonymous to a chunk. When the context is clear, we usually call a data/real packet a packet.
A data packet can be fairly large, such as 256 KB, and may need to be carried by multiple network-level packets. Our algorithms also use signaling/control packets, which are much smaller, e.g., under 400 bytes.}
arrive at the source according to a random process, which
will be distributed over the network to all the receivers, $V_s$.
The motivation for using a source model with dynamic arrivals is to account for the end-system bottleneck and timing variations in reading and transmitting locally stored data.
In some cases, the content may not be a static file or stored locally. The
model is general enough to cover realtime content, streaming video
with time-varying rate, or non-locally stored static data. Even if
the entire file is static and stored at the source, this source
model can still be useful. For instance, the data packets can be injected into the source node at a constant rate, which corresponds to a deterministic arrival process
with a constant arrival rate. Let $A_s(k)$ be the
number of packet arrivals on time slot $k$. Let us make the
following assumption on the arrival processes $\{A(k)\}$
throughout the paper, unless mentioned otherwise. Additional
assumptions may be added as needed.


\begin{assumption} \label{as:boundedstat}
For each source $s \in S$, $E[A_s(k) | x(k)] = \lambda_s$,
    and $E[(A_s(k))^2|x(k)] < K_1$ for some $0 < K_1 < \infty$, for all time $k$,
    where $x(k)$ represents the system state at time $k$.
\end{assumption}
The following are some remarks about Assumption AS \ref{as:boundedstat}.
\beit
\item
The system state, $x(k)$, will depend on the
specific settings of the two algorithms considered in this paper and will become clear later. It
usually includes all the queue sizes at time $k$, and possibly some additional
auxiliary variables. If the arrival process is independent of the past for each source $s$,
assumption AS \ref{as:boundedstat} can be stated without conditioning on $x(k)$. However, some non-IID arrival processes can lead to dependence between the current arrivals and the current queue sizes, in which case the statement with conditioning is more general than one without conditioning.


\item If the arrival process is independent of the past for each source $s$,
assumption AS \ref{as:boundedstat} can be stated without the conditioning.

\item $(\lambda_s)_{s \in S}$ could be the solution of a rate allocation
problem (see the discussion about the rate allocation and stability problems in Section \ref{sec:introduction}).

\eeit

\eat{

\item $B1$: For each source $s \in S$, $E[A_s(k)] = \lambda_s$;
    and $E[(A_s(k))^2] < K_1$ for some $K_1$, for all $k$.
    Furthermore, for every pair of
    sources $s$ and $s'$, the covariance
    $\text{Cov}(A_s(k), A_{s'}(k)) < K_2$ for some $K_2$,
    for all time $k$. That is, the second
    order statistics have uniform bounds.

\item $B2$:
    The processes $\{A_s(k)\}_k$ for different $s$ are
    independent from each other. For each fixed $s \in S$, $\{A_s(k)\}_k$ is a
    stationary, ergodic Markov chain.


\eeit

}

In this paper, we will present stable universal swarming
algorithms to distribute the packets to all the receivers.
For each $s \in S$, the packets will be transmitted along various
multicast distribution trees rooted at $s$ to the receivers in $V_s$.
A multicast tree in the multicast case corresponds to a path between a sender and a receiver in the unicast case.
Hence, using multiple multicast trees for a multicast session is analogous to data delivery using multiple paths between a sender and a receiver in the unicast case.


We will take Neely's definition of stability (\cite{Neely-IT06};
\cite{Neely03}, chapter 2) unless mentioned otherwise. For a single-queue
process $\{q(k)\}$, let us define the overflow function:
\be
\label{defnition:g_M}
g(M) = \limsup_{K \rightarrow \infty} \frac{1}{K} \sum_{k=1}^K P\{q(k) > M\}.
\ee
Roughly speaking, the overflow function $g(M)$ in (\ref{defnition:g_M}) defines the long-time average fraction of time when the queue size $q(k)$ is more than a chosen threshold $M$. In the stationary and ergodic case, it coincides with the stationary (and limiting) probability that the queue size exceeds $M$.
\begin{definition}
\label{definition:stability}
The single-queue process $\{q(k)\}$
is {\em stable} if $g(M) \rightarrow 0$ as $M \rightarrow \infty$. A network of
queues is stable if every queue is stable.
\end{definition}
With this definition of network stability, a sufficient condition
for network stability is:
Some Lyapunov function of the queues has a negative drift when any of the queues
becomes large enough \cite{Neely-IT06} \cite{Neely03} \cite{NML08} \cite{Neely2010}.
If with additional assumptions, the network queues
form an ergodic Markov chain, the same drift condition implies the chain
is positive recurrent, or equivalently, has a stationary distribution.

%

\subsection{Throughput Region}

For each source $s \in S$, let the set of candidate distribution
trees be denoted by $T_s$.
Throughout this paper, $T_s$ contains all
possible distribution trees rooted at the source $s$ unless
specified otherwise.
Let $T = \cup_{s \in S} T_s$. The trees
can be enumerated in an arbitrary order as $t_1, t_2, \cdots,
t_{|T|}$, where $|\cdot|$ denote the cardinality of a set.
Although $|T|$ is finite, it might be
very large.

The {\em throughput region} is defined as
\bea
\label{def:max_throughput_region}
\Lambda = \{\lambda \geq 0: \exists \alpha \geq 0
\mbox{ such that } \sum_{t \in T_s} \alpha_t = 1, \forall s \in S
\mbox{ and } \sum_{s \in S} \ \ \sum_{\{t \in T_s | e \in t\}} \alpha_t \lambda_s \leq c_e, \forall e \in E \}.
\eea
Here, $\alpha$ represents how the traffic from the sources
is split among the distribution trees.
The definition of $\Lambda$ says that a
source rate vector $\lambda$ is in $\Lambda$ if there exists a set of tree rates for each multicast session such
that the resulting total link data rate is no more than the link capacity
for any link.
Obviously, $\Lambda$ contains
the stability region, i.e., all $\lambda$ that
can be stabilized by some algorithms. This is so because,
for any non-negative mean rate vector
$\lambda \not\in \Lambda$, no matter how the traffic is split among
the distribution trees, there exists a link
$e$ such that the total arrival rate to $e$ is strictly
greater than its service rate.
Furthermore, this definition of the throughput region
allows the bandwidth bottleneck to be anywhere in
the network, at the access links or at the core.

We also define a {\em $\gamma$-reduced throughput region}
as $\frac{1}{\gamma}\Lambda$, where $\gamma \geq 1$.
By saying that the arrival rate vector $\lambda$
is {\em strictly inside the region $\frac{1}{\gamma} \Lambda$},
we mean that there exist some $\epsilon_0 > 0$
and a vector $\alpha \geq 0$ such that
$\sum_{t \in T_s} \alpha_t = 1, \forall s \in S$ and
$\sum_{s \in S} \sum_{\{t \in T_s | e \in t\}}
\alpha_t \lambda_s \leq \frac{1}{\gamma} c_e - \epsilon_0,
\forall e \in E$. This is equivalent to
\be
\label{ineq:capacity_region}
c_e \geq \gamma (\epsilon_0 + \sum_{s \in S}
\sum_{\{t \in T_s | e \in t\}} \alpha_t \lambda_s),
\ \ \forall e \in E.
\ee
Note that the region of rate vectors that are
strictly inside the region $\frac{1}{\gamma} \Lambda$ contains the
interior of $\frac{1}{\gamma} \Lambda$.
In Section \ref{sec:regulator}, we will show  that the interior of $\Lambda$ is stabilizable. That is,
for any rate vector $\lambda$ strictly inside the region $\Lambda$, there exists a scheduling algorithm such that the queues in the network are stable
under the algorithm.

\subsection{The Class of Algorithms: Time Sharing of Trees}

Each source has at least two possible approaches to use the
multicast trees. In one approach, the traffic from each source $s$
may be split according to some weights $(\alpha_t)_{t \in T_s}$
and transmitted simultaneously over the trees on every time slot.
Alternatively, the distribution can be done by time-sharing of the
trees. The algorithms in this paper follow the time-sharing
approach. On each time slot $k$, the source $s$ selects one
distribution tree from the set $T_s$, denoted by $t_s(k)$,
according to some tree-scheduling (tree-selection) scheme, and
transmits packets only over this tree on time slot $k$. The
time-sharing approach can emulate the first approach in the sense
that, when done properly, the fraction of time each distribution
tree is used over a long period of time can approximate any weight
vector $(\alpha_t)_{t \in T_s}$.

In addition to selecting the distribution tree $t_s(k)$ at each
time slot, an algorithm also needs to decide how many packets are
released to the tree. We will present two algorithms in the
following sections. The key question is what portion of the rate
region is stabilizable by each algorithm.



\section{Signaling, Source Traffic Regulation and
    $\gamma$-approximation Min-Cost Tree Scheduling}

\label{sec:regulator}

\subsection{Signaling Approach}



Stability analysis of a multi-hop network is often difficult
because the packets travel through the network hop-by-hop, instead
of being imposed directly to all links that they will traverse. As
a result, the arrival process to each internal link can be
difficult to describe. The frequently-used technique of network
signaling can be helpful.
In our case, on each time slot $k$, each source $s$ sends one
signaling packet to each node on the currently selected
tree $t_s(k)$. A signaling packet is one of the two types of control packets in the paper. It has two main functions. The first is to set up the multicast tree $t_s(k)$. A signaling packet contains a list of link IDs, which describe to the receiving node which of its outgoing links are part of the multicast tree. The second function is to inform the receiving node the intended source transmission rate on time slot $k$, i.e., the number of packets to be transmitted on time slot $k$.
To make the proofs for the main results easier, we make the following assumptions about all control packets, including the forward signaling packets and the feedback packets: The control packets are never lost and they arrive at their intended destinations within the same time slot on which they are first transmitted. Note that these assumptions are not crucial for either the theory or practice\footnote{
In actual operation, the two algorithms in the paper need not enforce these assumptions. The control packets can be lost or delayed. With straightforward minor modifications, such as using old information or postponing the algorithm execution until new information is available, the algorithms can cope with these conditions and are expected to be robust. They can achieve stability and throughput optimality as indicated by the theory. For better performance with respect to other metrics (e.g., data queue size, data delay, convergence speed), the algorithms can implement the following policy: The control packets are given higher priority at all the nodes than the data packets so that they experience minimal delay; the time slot size is chosen to be greater than the worst-case round-trip propagation time. With such a policy, the assumptions are expected to be satisfied most of the time.
In our simulation experiments (see Section \ref{sec:simulations}), the control packet delays are included. The algorithms converge fast and to the right values. On the theory side, there are good reasons to believe that the stability results in this paper still hold under the significantly relaxed assumptions: The network delays of the control packets are bounded and the number of consecutive control-packet losses is bounded.
The boundedness assumptions make the conditions of Corollary 1 in \cite{NML08} satisfied. The stability results would follow.
}. We do not make any assumptions on the data packets.

We will see that the rate information contained in a signaling packet is a very tight upper bound on the number of real packets released by the source on that time slot. To mark the the slight discrepancy, we use the term {\em virtual packets} and call the rate contained in the signaling packet the {\em rate of virtual packets} or the {\em virtual source rate}.
Consider a particular time slot $k$ and a particular internal link
$e$ on the selected distribution tree. The real packets issued by
the source on time slot $k$ will in general be delayed or buffered
at upstream hops and will not arrive at link $e$ until later.
However, via signaling, link $e$ knows how many virtual packets are
injected by the source and arrive at link $e$ on time slot $k$.
The cumulative number of arrived real packets at link $e$ must be no more than the
cumulative number of arrived virtual packets (via signaling packets).


One question is how many real packets
are to be released to the network on a time slot. One possibility
is that each source $s$ releases all the packets that arrive
during time slot $k$, i.e., $A_s(k)$. However, the uncontrolled
randomness of $A_s(k)$ causes difficulty in the stability
analysis, as we will see later. In our algorithm,
each source $s$ sets the number of real packets to be released
at the constant value $\lambda_s + \epsilon_1$ on every time slot
$k$, if that number of real packets is available. Every signaling packet from source $s$ contains the constant virtual packet rate $\lambda_s + \epsilon_1$.
Here, $\epsilon_1$ is a sufficiently small constant such that
$0 < |S| \epsilon_1 < \epsilon_0$. This guarantees the stability
of the source regulators, as we will see.

In the algorithm, each link $e$ updates a virtual queue,
denoted by $q_e(k)$.
\be
\label{eq:q_update_regulator}
q_e(k + 1) = [q_e(k) + \sum_{s \in S: e \in t_s(k)} (\lambda_s + \epsilon_1)
- c_e]_+.
\ee
$[\cdot]_+$ is the projection operation onto the non-negative domain. Note that the second term on the right hand side of (\ref{eq:q_update_regulator}) is the aggregate virtual data arrival rate from all the trees containing link $e$, which means the link capacities are shared by different trees.
Tree scheduling is based on the virtual queues instead of the real
queues.

\subsection{Source Traffic Regulation}


A regulator is placed at each source $s$ to ensure that on each time slot, source $s$ transmits no more than  $\lambda_s + \epsilon_1$ real packets.
A regulator is a traffic shaping device.
All the packets arriving at source $s$ first
enter a regulator queue. They will be released
to the network later in a controlled fashion.
On each time slot $k$,
let $D_s(k)$ denote the number of real packets released from the
regulator to the distribution tree $t_s(k)$,
and let $p_s(k)$ be the regulator queue size at source $s$.
The evolution of the regulator queue is given by
\be
\label{eq:regulator_update}
p_s(k + 1) = p_s(k) + A_s(k) - D_s(k),
\ee
where
\be
\label{eq:regulator_departure}
D_s(k) =
\left\{ \begin{array}{ll}
         \lambda_s + \epsilon_1 & \mbox{if $p_s(k) \geq \lambda_s + \epsilon_1 $};\\
        p_s(k) & \mbox{otherwise}.\end{array} \right.
\ee
Expressions (\ref{eq:regulator_departure}) ensures
that at most $\lambda_s + \epsilon_1$ real packets
are released on each time slot. Since
this departure rate is higher than
the mean packet arrival rate, stability of the regulator is guaranteed\footnote{
A packet may experience some delay at the
regulator before it is released to the network. The performance
analysis shows that this delay is bounded (in expectation) since all regulator queues are bounded. Our simulation results show that the delay and queue sizes can be made small even for very small $\epsilon_1$.}.
We will provide more details in the stability analysis.
Note that the traffic regulators are required only at the sources and they can be
implemented at the end-systems, i.e., the content servers.

\subsection{$\gamma$-Approximation Min-Cost Tree Scheduling}

We can interpret the virtual queue size $q_e$ as the cost of
link $e$. Then, the cost of a tree $t$ is $\sum_{e \in t} q_e$.
We propose the {\em $\gamma$-approximation min-cost tree scheduling} scheme:
On each time slot $k$ and for each source $s$, the selected
tree $t_s(k)$ satisfies
\be
\label{ineq:gamma_tree_regulator}
\sum_{e \in t_s(k)} q_e(k)
\leq \gamma \min_{t \in T_s} \sum_{e \in t} q_e(k),
\ee
where $\gamma \geq 1$.
If there are multiple trees satisfying (\ref{ineq:gamma_tree_regulator}),
the tie is broken arbitrarily.

The rationale for this tree-scheduling scheme is straightforward.
When $\gamma = 1$, the tree-scheduling scheme solves
the min-cost Steiner tree problem, which is
NP-hard. But, the min-cost Steiner tree problem has
approximation solutions, which we can use.
In \cite{ChaChe99}, a family of approximation algorithms for the directed
Steiner tree problem is proposed, which achieves an $O(\log^2 N)$ approximation
ratio in quasi-polynomial time, where $N$ is the number of receivers.
It will be proven in the following stability analysis that,
if we are able to find
the minimum-cost Steiner tree on each time slot, we can stabilize the network
for the interior of the entire throughput region, $\Lambda$;
if we adopt the $\gamma$-approximated min-cost tree scheduling, we
can stabilize the network
for the interior of $\frac{1}{\gamma} \Lambda$.

The link costs (i.e., the virtual queue sizes) are carried to the multicast sources by the second type of control packets - the feedback packets. On each time slot, a network node sends to each source one feedback packet, which contains the costs of the node's outgoing links.

\subsection{Stability Analysis}

The stability analysis is based on the drift analysis
of Lyapunov functions.

\subsubsection{Stability of the Regulators}

Define a Lyapunov function of the regulator queues
$p$ as
\be
\label{eq:lyapunov_p_regulator}
L_1(p) = \sum_{s \in S} p_s^2.
\ee
\begin{lemma}
\label{lemma:lyapunov_drift_p_regulator}
There exists some positive constant $0 < M_1 < \infty$ such that
for every time slot $k$ and the regulator backlog vector $p(k)$, the
Lyapunov drift satisfies
\be
\label{ineq:lyapunov_drift_p_regulator}
E[L_1(p(k+1)) - L_1(p(k)) | p(k)]
\leq - \epsilon_1 \sum_{s \in S} p_s(k),
\ee
if $\sum_{s \in S} p_s(k) \geq \frac{M_1}{\epsilon_1}$.
\end{lemma}
\IEEEproof{
This is because the mean arrival rate is strictly less
than the mean service rate provided the regulator has
sufficient packets. The proof is standard and we omit the details.
}

\subsubsection{Stability of the Virtual Queues}

Define a Lyapunov function of the virtual queue backlog vector
$q$ as
\be
\label{eq:lyapunov_q_regulator}
L_2(q) = \sum_{e \in E} q_e^2.
\ee

Let $t(k) = (t_s(k))_{s \in S}$ be the vector of the chosen distribution
trees at time $k$.
We allow randomness in determining $t(k)$. For instance,
when there are multiple trees satisfying
(\ref{ineq:gamma_tree_regulator}), the tie can be broken randomly.
\begin{lemma}
\label{lemma:lyapunov_drift_q_regulator}
If the mean arrival rate vector $\lambda$ is strictly
inside the region $\frac{1}{\gamma}\Lambda$, then,
there exist some positive constants $0 < M_2 < \infty$ and $\epsilon$ for
all sample paths of $\{t(k)\}_{k}$ such that,
for every time slot $k$ and virtual queue backlog vector $q(k)$,
the Lyapunov drift satisfies
\be
\label{ineq:lyapunov_drift_q_regulator}
L_2(q(k+1)) - L_2(q(k)) \leq M_2 - 2 \epsilon \sum_{e \in E} q_e(k),
\ee
where $\epsilon = \gamma \epsilon_0 - |S| \epsilon_1 > 0$.
\end{lemma}
\IEEEproof{
Under any sample path of the process $\{t(k)\}_{k}$,
(\ref{ineq:lyapunov_drift_q_regulator}) holds due to
the tree-selection algorithm (\ref{ineq:gamma_tree_regulator})
and the definition of the $\gamma$-reduced
throughput region (\ref{ineq:capacity_region}).
The detailed proof is omitted for brevity.
}

Hence, when $\sum_{e \in E} q_e(k) \geq \frac{M_2}{\epsilon}$,
the Lyapunov function has a negative drift under all sample paths of $\{t(k)\}_{k}$.
\be
L_2(q(k+1)) - L_2(q(k)) \leq - \epsilon \sum_{e \in E} q_e(k).
\ee

\begin{corollary}
\label{corollary:q_bound_regulator}
For each link $e$, there exists a sufficiently
large constant $M_e < \infty$
such that $q_e(k) \leq M_e$.
\end{corollary}
\IEEEproof{
From Lemma \ref{lemma:lyapunov_drift_q_regulator},
under any sample path of $\{t(k)\}_{k}$,
$q_e(k)$ is uniformly bounded from above.
Hence, there exists a sufficiently large constant $M_e < \infty$
such that $q_e(k) \leq M_e$.
}

\noindent {\bf Remark}: The chosen deterministic release rates of the virtual packets
guarantee that the virtual queues are bounded. This is an important
fact for proving the stability of the real queues.
If the sources signal the actual numbers of real packet arrivals on each time slot, which are random, the
virtual queues may be stable but are not guaranteed to be bounded.

\subsubsection{Stability of the Real Queues}

For convenience, let us assume each real packet remembers its distribution
tree. This way, the nodes on the tree know when to duplicate the packet.
Moreover, each packet at any link also has
an unambiguous {\em hop count}, which
is the hop count on its tree path from the source to the current link. With this setup,
we can assume to use the following queueing discipline for the
real queues.
\begin{assumption} \label{as:priorityqueue}
At each link $e$, a packet with a smaller hop count has
priority over any packet with a larger hop count.
\end{assumption}

First, we will show some properties of the real packet
arrival rates to the intermediate links.
Define an indicator function $I(e, t)$, where $e$ is a link and $t$ is a tree.
\be
I(e, t) =
\left\{ \begin{array}{ll}
         1 & \mbox{if $e \in t$};\\
        0 & \mbox{otherwise}.\end{array} \right. \nonumber
\ee
\begin{lemma}
\label{lemma:link_arrival_bound_regulator}
For any link $e \in E$, there exists a constant $0 < M_e < \infty$
such that for any $k_0$ and $k$ with $k_0 \leq k$,
\be
\label{ineq:link_arrival_bound_regulator}
\sum_{u = k_0}^k \sum_{s \in S} D_s(u) I(e, t_s(u))
\leq (k - k_0 + 1) c_e + M_e.
\ee
\end{lemma}
\IEEEproof{
According to (\ref{eq:q_update_regulator}),
\be
\label{ineq:link_arrival_proof_1}
q_e(k + 1) \geq q_e(k) + \sum_{s \in S}(\lambda_s + \epsilon_1) I(e, t_s(k))
- c_e,
\forall e \in E.
\ee
By summing (\ref{ineq:link_arrival_proof_1}) over time slots,
the proof is straightforward and is omitted.
}

Let $Q_e(k)$ denote the real queue backlog of link $e$ at time slot $k$.
We can show by induction that under the prioritized queueing strategy
in AS \ref{as:priorityqueue},
the real queue backlogs are bounded. The proof is adapted from \cite{LS06A}.

\begin{theorem}
\label{theorem:Q_bound_regulator}
With the additional assumption AS \ref{as:priorityqueue},
if the mean arrival rate vector $\lambda$ is strictly
inside the region $\frac{1}{\gamma}\Lambda$,
the real queue backlogs are bounded. I.e., there exists some constant
$0 < M' < \infty$ such that
\be
\label{ineq:Q_bound_regulator}
Q_e(k) \leq M', \ \ \ \forall k, \ \forall e \in E.
\ee
\end{theorem}
\IEEEproof{
See Appendix A.
}

\begin{theorem}
\label{thm:stability_regulator}
With the additional assumption AS \ref{as:priorityqueue},
if the mean arrival rate vector $\lambda$ is strictly
inside the region $\frac{1}{\gamma}\Lambda$, $\gamma \geq 1$,
the $\gamma$-approximation min-cost tree scheduling
scheme stabilizes the network.
\end{theorem}
\IEEEproof{
From Lemma \ref{lemma:lyapunov_drift_p_regulator},
Lemma \ref{lemma:lyapunov_drift_q_regulator}
(or Corollary \ref{corollary:q_bound_regulator}) and
Theorem \ref{theorem:Q_bound_regulator}, the regulator queues have
negative drifts, the virtual queues and the real queues in the
network are all bounded.
}

The reduction factor $1/\gamma$ in Theorem
\ref{thm:stability_regulator} is a lower bound for the worst case. In practice, the actual reduction factor for a given network and a specific tree-scheduling scheme may be much larger. For instance, we use a sub-optimal tree-scheduling scheme and some ISP networks for the simulation results in Section \ref{sec:simulations}, and the algorithm can achieve nearly the full throughput region.



\section{Randomized Tree Scheduling}
\label{sec:without_regulator}

Theorem \ref{thm:stability_regulator} implies that the interior of
the throughput region $\Lambda$ can be stabilized, provided one
can solve the hard min-cost Steiner tree problem. If approximation
algorithms are used for the Steiner tree problem, Theorem
\ref{thm:stability_regulator} says that a reduced rate region is
stabilizable.
In this section, we will continue to cope with the hard Steiner
tree problem. Instead of approximation algorithms, we will
consider an algorithm that randomly samples the trees at each time
slot. Selecting trees by random sampling is attractive in practice
since the algorithms for doing this tend to be simple and fast.
Some practical systems such as BitTorrent \cite{BitTorrent}
already use variants of random sampling.

Our main concern is whether the tree-sampling approach has any
performance guarantee with respect to stability. We conjecture it
does.
We will show important steps that may eventually lead to the
conclusion that, in contrast to the case with approximation
algorithms, the entire interior of $\Lambda$ is stabilizable. The
development and analysis of the algorithm are in part based on
\cite{Tassiulas98}.

\subsection{Signaling}

In this algorithm, the sources still signal the links about the incoming traffic,
but they are not regulated. Specifically,
the number of virtual packets signaled by source $s$ on every time slot $k$
is $A_s(k)$ instead of $\lambda_s + \epsilon_1$.
For each $e \in E$, the evolution of the virtual queue, $q_e(k)$, is
\be
\label{eq:q_update}
q_e(k + 1) = [q_e(k) + \sum_{s \in S: e \in t_s(k)} A_s(k) - (c_e - \epsilon_2)]_+,
\ee
where $0 < \epsilon_2 < \epsilon_0$.
From (\ref{eq:q_update}), the virtual queue is serviced at less than the full service capacity.
We will see the reason in the stability analysis (see Corollary \ref{corollary:traffic_intensity} and the remark after it).


\subsection{Randomized Tree Scheduling}

Let $\tau_s(q)$ denote the min-cost tree for source $s$
with respect to the link cost vector $q$.
We have,
\be
\label{eq:min_tree}
\sum_{e \in \tau_s(q)} q_e
= \min_{t \in T_s} \sum_{e \in t} q_e.
\ee
If multiple min-cost trees exist, an arbitrary one is chosen.

The algorithm has two stages: {\em pick} and {\em compare}. In the pick stage,
each source uses some randomized
algorithm to pick a tree, with the requirement that there is a positive probability
to pick a min-cost tree.
More specifically, let $\hat{t}(k)=(\hat{t}_s(k))_{s \in S}$
be the trees picked by the randomized algorithm
on time slot $k$.
The following condition is satisfied for some $\delta > 0$,
\be
\label{eq:random_min_tree}
P \{\sum_{e \in \hat{t}_s(k)} q_e(k) = \sum_{e \in \tau_s(q(k))} q_e(k),
\forall s \in S \} \geq \delta.
\ee

In the compare stage, the cost of the tree just picked is compared with
the cost of the selected tree on the previous time slot, with respect to the current link cost
vector. The picked tree is selected only if it has a lower cost.
This ensures that the tree that ends up being selected is better than the previously selected tree.
Recall that $t_s(k)$ is the scheduled tree at time $k$. The compare stage yields a selected tree
that satisfies the following: For any source $s \in S$,
\begin{align}
& t_s(k)
= \left\{ \begin{array}{ll}
        \hat{t}_s(k) & \mbox{if $\sum_{e \in \hat{t}_s(k)} q_e(k) \leq \sum_{e \in t_s(k-1)} q_e(k)$};\\
        t_s(k-1) & \mbox{otherwise}.\end{array} \right. \label{eq:sampled_min_tree}
\end{align}

There are many possible randomized selection algorithms that satisfy
(\ref{eq:random_min_tree}) and (\ref{eq:sampled_min_tree}).
For instance, one
algorithm might be to modify the current tree by randomly adding or deleting
edges until a new multicast tree is found.
The selection of the edges can be biased toward lower-cost ones for addition
and higher-cost ones for deletion.
In this paper,
we will not dwell on finding specific algorithms
but will focus on the stability issue of the whole algorithm class.

\subsection{Stability Analysis}

We will show that, if the mean arrival rate vector $\lambda$ is
strictly inside the throughput region $\Lambda$, the
randomized tree-scheduling scheme is able to stabilize all the
virtual queues. With additional assumptions, the cumulative
arrival of the real packets by any time slot is strictly less than
the accumulation of the link service rate for every link.


\subsubsection{Stability of the Virtual Queues}

The virtual queue sizes $q(k)$ are considered as the link costs. Let $t(k)$ be
the vector of chosen trees.
Define a Lyapunov function of $x = (q, t)$:
\be
L(x) = L_1(x) + L_2(x), \nonumber
\ee
where
\be
L_1(x) = \sum_{e \in E} q_e^2 \nonumber, \ \ \
L_2(x)
= (\sum_{s \in S} \lambda_s (\sum_{e \in t_s} q_e
- \sum_{e \in \tau_s(q)} q_e))^2.
\nonumber
\ee

The proofs for the following three main lemmas parallel the development
in \cite{Tassiulas98}, although the details are different and technical.
We omit them for brevity.

\begin{lemma}
\label{lemma:L1_random}
If the mean arrival rate vector $\lambda$ is strictly inside the throughput
region $\Lambda$,
there exist some positive constants $M_1$ and $\epsilon$ such that
\begin{align}
\label{ineq:L1_random}
& E[L_1(x(k+1)) - L_1(x(k)) | x(k)]
\leq M_1 + 2 \sqrt{L_2(x(k))} - 2 \epsilon \sum_{e \in E} q_e(k).
\end{align}
\end{lemma}

\begin{lemma}
\label{lemma:L2_random}
If the arrival rate vector $\lambda$ is strictly inside the throughput
region $\Lambda$,
there exist some positive constants $M_2$ and $M_3$ such that
\begin{align}
\label{ineq:L2_random}
& E[L_2(x(k+1)) - L_2(x(k)) | x(k)]
\leq M_2 + M_3 \sqrt{L_2(x(k))} - \delta L_2(x(k)).
\end{align}
\end{lemma}

\begin{lemma}
\label{lemma:lyapunov_drift_random}
If the arrival rate vector $\lambda$ is strictly inside the throughput
region $\Lambda$,
there exist some positive constants $M < \infty$ and $\epsilon$ such that,
if $L(x(k)) \geq M$,
\be
\label{ineq:lyapunov_drift_random}
E[L(x(k+1)) - L(x(k)) | x(k)]
\leq - \epsilon \sqrt{L_1(x(k))}.
\ee
\end{lemma}

\begin{theorem}
\label{thm:stability_q_random}
If the mean arrival rate vector $\lambda$ is strictly
inside the throughput region $\Lambda$,
the randomized tree scheduling scheme stabilizes the virtual queues.
\end{theorem}
\IEEEproof{
This is a corollary from Lemma \ref{lemma:lyapunov_drift_random}.
}


\subsubsection{Stability of the Real Queues}

We have partial results about the stability of the real queues
under additional conditions. We assume the following in this
subsection.

\begin{assumption} \label{as:formarkov}
    The processes $\{A_s(k)\}_k$ for different $s$ are
    independent from each other. For each $s \in S$, $\{A_s(k)\}_k$ is
    IID. At every time $k$, there is
    a nonzero probability that no packet arrives
    at the sources, i.e.,
    $\text{P}\{A_s(k) = 0, \forall s \in S\} > 0$.
\end{assumption}

We will show that for any link $e$, its average traffic intensity (load),
$\rho_e$, satisfies
$\rho_e < 1$, where $\rho_e$ is the ratio of the average
packet arrival rate and the link rate. First,
stronger stability conclusions can be said about the virtual queues.

\begin{theorem}
\label{theorem:q_ergodicity_regularity}
Suppose the mean arrival rate vector $\lambda$ is strictly
inside the throughput region $\Lambda$, and
assumptions AS \ref{as:boundedstat} and AS \ref{as:formarkov} hold.
\beit
\item The process $\{q(k), t(k) \}_{k=0}^{\infty}$ is an aperiodic and
    irreducible Markov chain with a stationary
    distribution. Moreover, let $\hat{q}$ be the virtual queues under the
    stationary distribution. Then,
    $E[\hat{q}_e] < \infty$.
\item The strong law of large numbers holds: For each
    initial condition, and for all $e \in E$,
\be
\label{eq:q_law_large_number}
\lim_{k \rightarrow \infty} \frac{\sum_{u = 0}^k q_e(u)}{k + 1}
= E[\hat{q}_e], \mbox{ almost surely.}
\ee
\item The mean ergodic theorem holds: For each initial condition,
and for all $e \in E$,
\be
\label{eq:q_mean_ergodic}
\lim_{k \rightarrow \infty} E [q_e(k)] = E[\hat{q}_e].
\ee
\eeit
\end{theorem}
\IEEEproof{
See Appendix A.
}

\begin{theorem}
\label{theorem:cumulative_arrival}
For any link $e \in E$,
\be
\label{ineq:cumulative_arrival_strong_law}
\limsup_{k \rightarrow \infty}
\frac{1}{k+1}\sum_{u = 0}^k \sum_{s \in S} A_s(u)I(e, t_s(u))
\leq c_e - \epsilon_2,
\ee
\be
\label{ineq:cumulative_arrival_mean_ergodic}
\limsup_{k \rightarrow \infty}
E[\frac{1}{k+1}\sum_{u = 0}^k \sum_{s \in S} A_s(u)I(e, t_s(u))]
\leq c_e - \epsilon_2.
\ee
\end{theorem}
\IEEEproof{
See Appendix A.
}

Recall that $Q_e(k)$ denotes the real queue backlog of link $e$ at time slot $k$.
Next, we show that the process $\{Q_e(k)\}$ is {\em rate stable}
for all links, where the defintion of rate stability is given
as in \cite{Neely2010}.
\begin{corollary}
\label{corollary:Q_rate_stable}
Suppose the mean arrival rate vector $\lambda$ is strictly
inside the throughput region $\Lambda$, and
assumptions AS \ref{as:boundedstat} and AS \ref{as:formarkov} hold.
For any link $e \in E$, the process
$\{Q_e(k)\}$ is rate stable, i.e.,
\be
\lim_{k \rightarrow \infty} \frac{Q_e(k)}{k} = 0 \ \
\mbox {with probability } 1.
\nonumber
\ee
\end{corollary}
\IEEEproof{
See Appendix A. \\
}
Rate stability implies that the long-term average
rates of arrivals and departures are identical for each queue,
and is weaker than the
stability definition of the queue backlog being bounded.

\begin{corollary}
\label{corollary:traffic_intensity}
For any link $e \in E$, the average traffic intensity (or load)
$\rho_e < 1$, where $\rho_e$ is defined as
\be
\nonumber
\rho_e =
\limsup_{k \rightarrow \infty}
\frac{\sum_{u = 0}^k a_e(u)}
{(k+1) c_e},
\ee
where $a_e(u)$ is the number of real packets arriving at link $e$ at time $u$.
\end{corollary}
\IEEEproof{
For any time slot $k$,
the number of cumulative arrivals at the real queue is no
more than the number of cumulative arrivals at the virtual queue,
i.e.,
\be
\nonumber
\sum_{u = 0}^k a_e(u)
\leq
\sum_{u = 0}^k \sum_{s \in S} A_s(u)I(e, t_s(u)).
\ee
Then, $\rho_e < 1$ follows from Theorem \ref{theorem:cumulative_arrival}.
}

\noindent {\bf Remark:} The service rate of the virtual queue of link $e$,
which is $c_e - \epsilon_2$, guarantees
$\rho_e < 1$.

Under the randomized tree scheduling scheme,
the virtual queues are stable, the real queue
processes $\{Q_e(k)\}$ are rate stable, and the real traffic intensity satisfies
$\rho_e < 1$ for every link $e$. But, we do not know whether
the real queues are stable in the sense of Definition \ref{definition:stability}.
We expect that in practice, they are almost always stable.
We suspect that under more assumptions on the traffic arrival processes and
the queueing discipline, the real queues
can be proven to be stable.


\section{Simulation Results and Evaluations}
\label{sec:simulations}

In this section, we present illustrative examples from simulation experiments that support the stability analysis of the algorithms. We will also evaluate the control overhead of the algorithms.

Since the algorithms are based exclusively on the information contained in the control packets (including the forward signaling and reverse feedback packets), we trace the behavior of the control packets carefully with event-driven simulation at the packet level. Link propagation delays and transmission delays for the control packets are included in the simulation. The control packets are routed on the shortest path, measured by the hop count. At each link, the control packets are transmitted at a higher priority than the data packets and, if needed, are stored in a high-priority queue. When a control packet arrives at its intended destination, an event will be triggered to update the virtual queues or the network costs, depending on whether it is a forward signaling or a feedback packet. The rest of the algorithm operations take place on time slot boundaries. On each time slot, a source sends one signaling packet to each node on the currently selected multicast tree to set up the tree and to inform the virtual source rate; it transmits data to the tree in the amount decided by the algorithms; it also computes a new tree according to the network costs that it currently knows, and the new tree will be set up and used on the next time slot. On each time slot, a network node sends at most one set of link costs (of its outgoing links) to each of the multicast sources.

To evaluate the stability of the real queues, we also need to track the sizes of the real data queues. In the interest of reducing simulation time, we trace the real data at the burst level instead of the packet level. Specifically, for each link, the simulator computes the amount of data it can transmit in the time slot, which is the difference of
the link capacity and the amount of control packets transmitted
during that time slot. Then, the burst of data is pushed to the next hop. Although there is a slight degree of inaccuracy in the simulated queue sizes, the outcome of whether or not the queues are stable is not altered by the burst-level simulation for data.

We simulate our algorithms over two commercial ISP network topologies obtained from the Rocketfuel project \cite{RocketFuel}. The first one consists of $41$ nodes and $136$ links; the second one consists of $295$ nodes and $1086$ links. For each network, we assume the link capacities are exclusively allocated to the content distribution service.
We assume there is a single distribution session.

On the smaller network with $41$ nodes and $136$ links, we select one node as the source and $20$ nodes as the receivers. All other out-of-session nodes may be used as helper nodes. We assign $1$ Gbps link capacity to all the links except some critical links. By critical links, we mean the links that become bottleneck easily if they do not have sufficient capacity. We assign $5$ Gbps link capacity to each of the critical links. There are exogenous random arrivals of packets to the source. We assume that the number of packet arrivals on each time slot is a Poisson random variable and that the arrivals on different time slots are IID. The size of each data packet (chunk) is chosen to be $256$ KB. Since the time slot size is equal to $1$ second, the mean of the Poisson distribution is equal to $\lambda_s / (256 \times 8 \times 1000)$ and the unit is in packets. As an example, for the source arrival rate $\lambda_s = 1990$ Mbps, the mean number of arrivals is about $972$ packets. The standard deviation is about $31.2$ packets. The Poisson distribution is widely used to capture the total effect of many small disturbances when the outcome is non-negative and integer-valued.
The maximum achievable session rate is $2$ Gbps, which is obtained by running the subgradient algorithm introduced in \cite{ZCX09-1a}.
The control packet size is under 400 bytes for our experiments. The time slot duration is 1 second.
We have done experiments with different link propagation delays: $20$ ms, $50$ ms, $80$ ms, and $100$ ms. These cases have similar performance results.
The 1-second time slot size is the relevant delay that determines the algorithm performance. Hence, we will only present the results for the case of $100$ ms propagation delay at each link.
We vary the mean arrival rate $\lambda_s$ to see whether the algorithms can achieve network stability if the rate is below the maximum achievable session rate.

We also conducted experiments with other traffic models, such as truncated Pareto distributions, which have very large variances, and other distributions with very small variances. In addition, we conducted experiments where multiple multicast sessions exist simultaneously in the network. The results for these cases do not show much more insight than what we will subsequently present, and for brevity, they are not reported in the paper.


\subsection{Algorithm Using Source Traffic Regulation and Approximate Min-Cost Tree Scheduling}

In this subsection, we show the performance of the algorithm introduced in Section \ref{sec:regulator}.
For the approximate tree selection algorithm, we use the algorithm by Charikar \emph{et. al} with tree level $2$, as proposed in \cite{ChaChe99}.
A regulator queue is maintained only at the source.
In the simulation, we set $\epsilon_1 = 1$ packet per second or $2.048$ Mbps.
Our main concern is whether the regulator and real queues are bounded if the mean arrival rate to the source is below the maximum achievable session rate.

\begin{figure}[t]
\begin{center}
\begin{minipage}{3.2in}
\begin{center}
\includegraphics[width=3.2in]{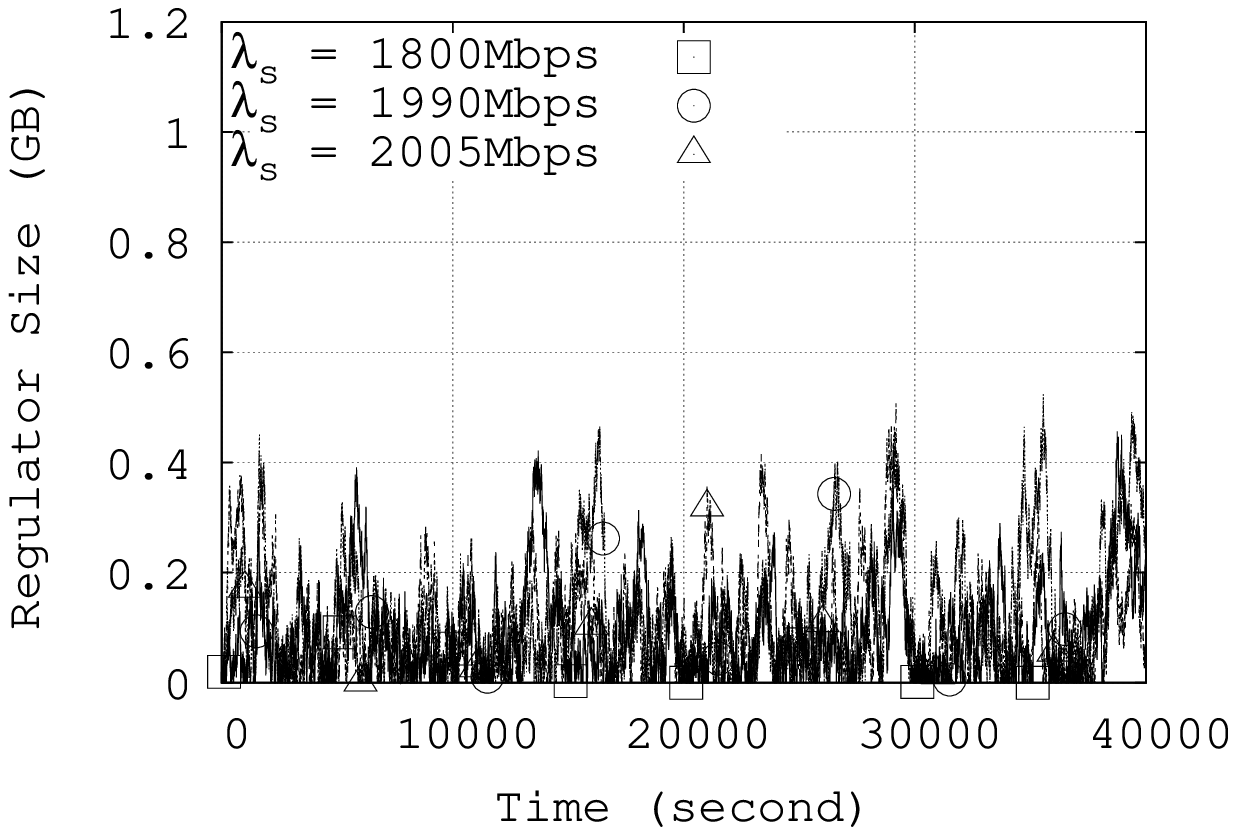} \\ (a)
\end{center}
\end{minipage}
\begin{minipage}{3.2in}
\begin{center}
\includegraphics[width=3.2in]{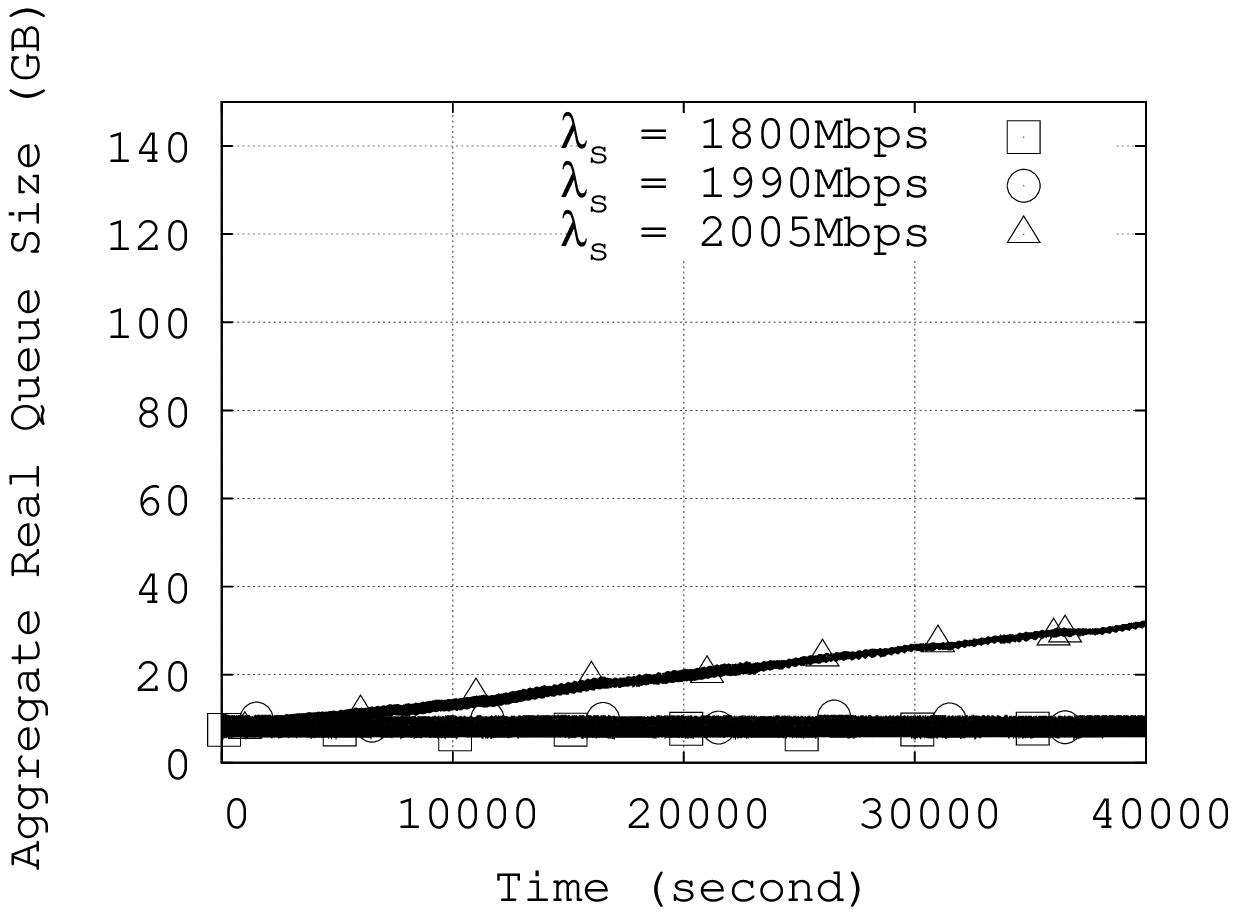} \\ (b)
\end{center}
\end{minipage}
\end{center}
\caption{Peformance of the first algorithm; (a) regulator queue size when $\epsilon_1 = 1$ packet per second (or $2.048$ Mbps);
(b) aggregate real queue sizes.} \label{fig:StabilityFirstAlgorithm}
\end{figure}

Fig. \ref{fig:StabilityFirstAlgorithm} shows that the network queues remain stable if $\lambda_s$ is below the maximum achievable session rate ($2$ Gbps) even when $\lambda_s$ is quite close to the maximum session rate. when $\lambda_s$ exceeds the maximum achievable session rate, the total network queue size grows without a bound, which means some of the queues are unstable.

Fig. \ref{fig:StabilityFirstAlgorithm} (a) shows that the regulator queue size is bounded
and is less than $500$ MB even when $\lambda_s$ is greater than the maximum achievable session rate. The reason is that the regulator queue is a simple single-server queue with a deterministic service rate and the algorithm sets
the service rate to be slightly greater than $\lambda_s$, by $\epsilon_1$ as given in (\ref{eq:regulator_departure}). In the experiments here, $\epsilon_1$ is very small relative to the traffic arrival rate and the traffic load is extremely heavy. For instance, when $\lambda_s = 1990$ Mbps, the traffic load (intensity) to the regulator queue is $0.999$. Even under such heavy load, the regulator queue size is not very large.
It can be made much smaller when $\epsilon_1$ is increased.
Fig. \ref{fig:StabilityFirstAlgorithm} (b) shows that, eventually, the aggregate real queue size over all the queues in the network is under $11$ GB when the arrival rate ($\lambda_s = 1990$ Mbps) is slightly below the maximum achievable rate; that yields an average queue size of 80.9 MB at each link. The largest queue size at a link is about $625$ MB under that arrival rate\footnote{In this section, the maximum queue sizes are what we observed during a long simulation run.}. The queue sizes can be much smaller when the arrival rate is lower. These queue size values can be compared with the bandwidth-delay product, which is $625$ MB for a critical link and $125$ MB for a non-critical link.
Hence, using a $1$ GB buffer at each link is more than sufficient for this test case.
Since the source sends data packets at a rate of $1990$ Mbps and the packets need to be duplicated to $20$ receivers, about $5$ GB data must flow through the network every second. The $11$ GB data stored in all the queues is about twice of that amount. Hence, the amount of queued data is reasonably small.

\begin{figure}[t]
\begin{center}
\includegraphics[width=3.2in]{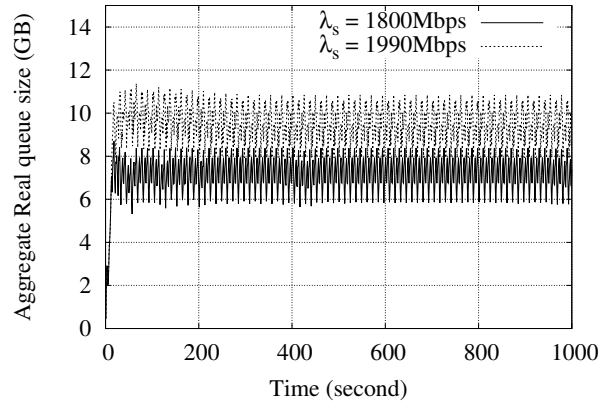}
\end{center}
\caption{Aggregate real queue sizes of the first algorithm on time slots $0$ to $1000$.} \label{fig:queueFirstAlgorithmzoom}
\end{figure}

It is important to point out that the eventual queue sizes shown in Fig. \ref{fig:StabilityFirstAlgorithm} are mostly determined by the transient phase of the algorithm, which is the phase at the beginning of the algorithm operation before the time-average rates approach the optimum. The queues build up at this phase because the algorithm hasn't found the right transmission rates yet. Once the time-average rates approach the optimum, the queues stop growing but oscillate around some values (see Fig. \ref{fig:queueFirstAlgorithmzoom}, which shows the aggregate queue sizes on time slots $0$ to $1000$). The oscillation is due to a feature of the algorithm, which is that the multicast session hops among different multicast trees even in the steady state. A consequence is that some of the links can be temporarily overloaded. From the simulation results, we see that the magnitude of the oscillation can be much smaller than the queue size itself. The oscillation of the aggregate queue size is less than $3$ GB when $\lambda_s = 1990$ Mbps, which yields an average of $22.1$ MB per link. When the multicast sessions are long-lasting and the network topology and link bandwidth are unchanging, it is enough to decide the buffer sizes based on the steady-state queue behavior. In the cases of Fig. \ref{fig:StabilityFirstAlgorithm}, we see that the buffer requirement is much smaller than the bandwidth-delay products.

The good queue-size performance can be explained by two observations. First, by using multiple multicast trees, the excess packets are spread out and queued all over the networks. Second, the algorithm converges reasonably fast. Fig. \ref{fig:receiving-rate} (a) shows that, for each experiment where the arrival rate is below the maximum achievable session rate, the average data receiving rate per receiver reaches more than 90\% of the arrival rate at around $100$ to $200$ seconds, and it reaches nearly 100\% of the arrival rate after $1000$ seconds. The average receiving rate is also a time average, which has long-time memory. Judging by Fig. \ref{fig:queueFirstAlgorithmzoom}, the queues of the system approach the steady state very quickly, within $50$ time slots. The instantaneous receiving rates must have reached the arrival rate within the same time frame.

\subsection{Algorithm Using Randomized Tree Scheduling}

In this subsection, we show the stability of the algorithm introduced in Section \ref{sec:without_regulator}.
For the randomized tree selection algorithm, we let each
link be selected as an edge on the random tree
with a probability inversely proportional to its
virtual queue size. The idea is to reduce the chance of selecting links with large virtual queues. Once a link is selected to be on the tree,
all links that will lead to a loop with the selected links are removed from the candidate list. The candidate links are scanned repeatedly in
a breadth-first order, starting from the source, until all the receivers are connected.
How the random tree is selected will not affect the stability result as long as the condition in (\ref{eq:random_min_tree}) is satisfied. However, the choice of the tree affects other aspects of performance, such as the queue sizes. When the queue sizes are considered in addition to throughput, what can be considered as good choices for the random tree remains an open question. Interested readers may refer to related literature in randomized link scheduling algorithms for wireless networks \cite{LR06, Ra09, Jiang10, NTS10}.

\begin{figure}[t]
\begin{center}
\begin{minipage}{3.2in}
\begin{center}
\includegraphics[width=3.2in]{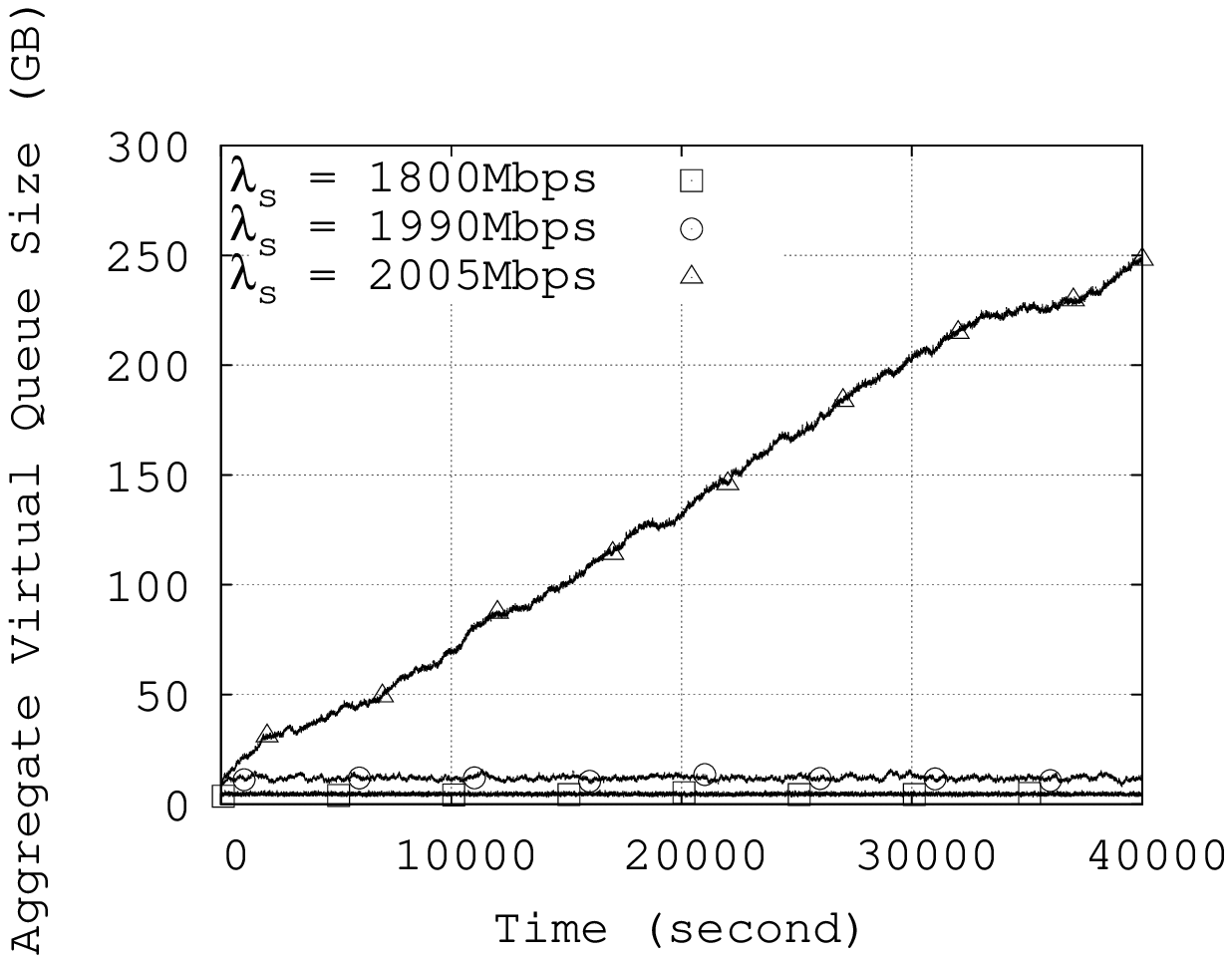} \\ (a)
\end{center}
\end{minipage}
\begin{minipage}{3.2in}
\begin{center}
\includegraphics[width=3.2in]{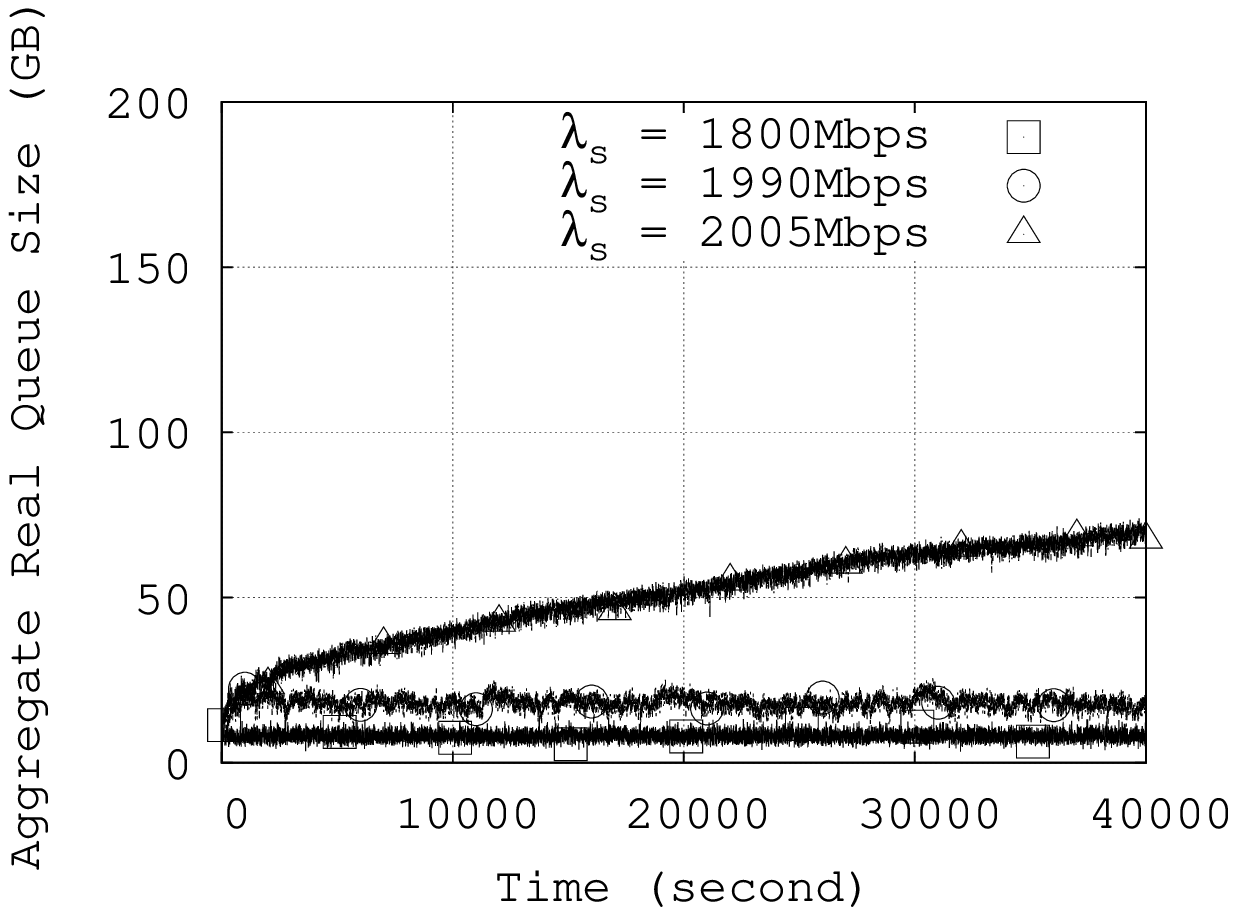} \\ (b)
\end{center}
\end{minipage}
\end{center}
\caption{The stability of the second algorithm;
(a) aggregate virtual queue sizes;
(b) aggregate real queue sizes.}
\label{fig:StabilitySecondAlgorithm}
\end{figure}

\begin{figure}[t]
\begin{center}
\begin{minipage}{3.2in}
\begin{center}
\includegraphics[width=3.2in]{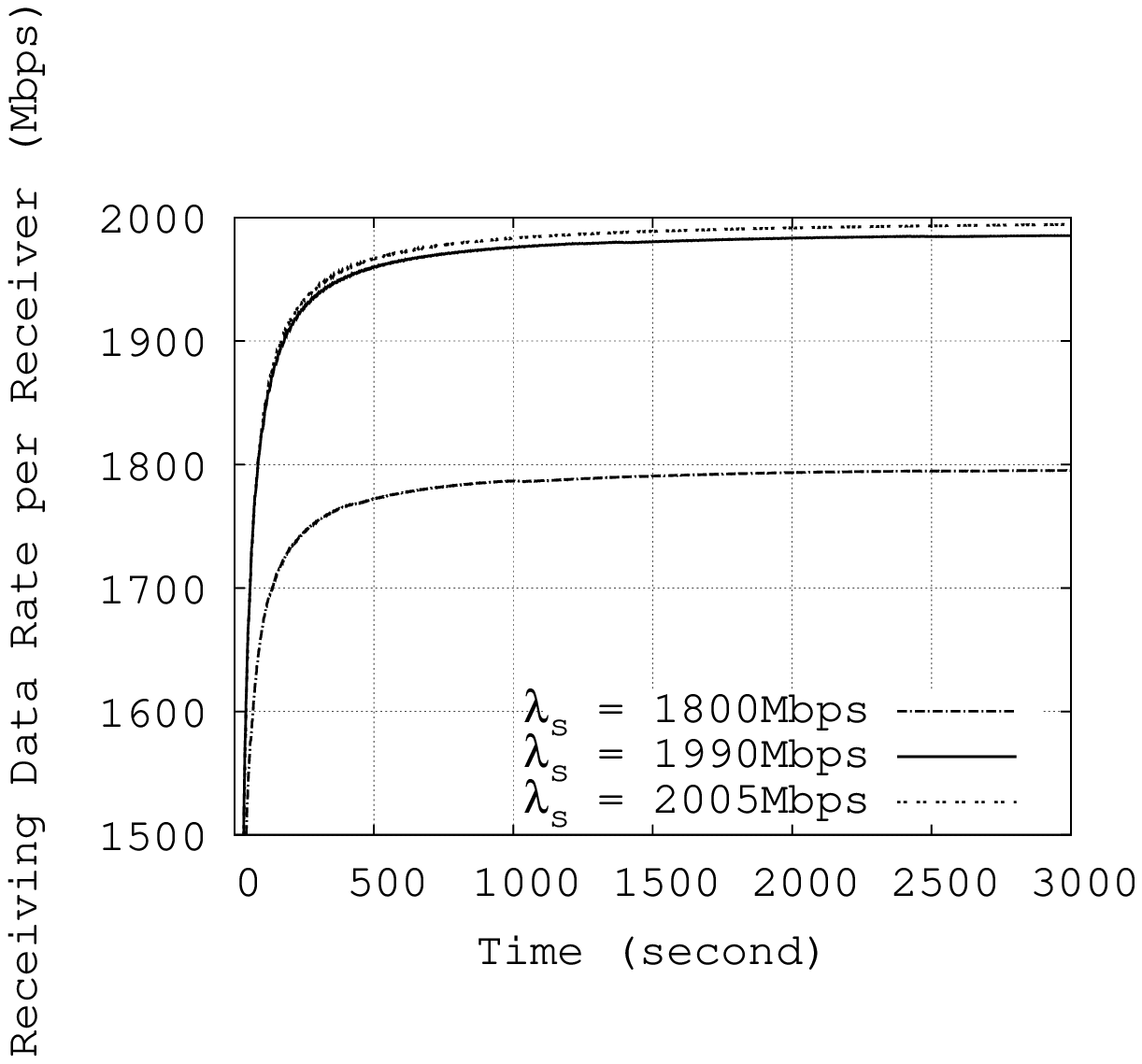} \\ (a)
\end{center}
\end{minipage}
\begin{minipage}{3.2in}
\begin{center}
\includegraphics[width=3.2in]{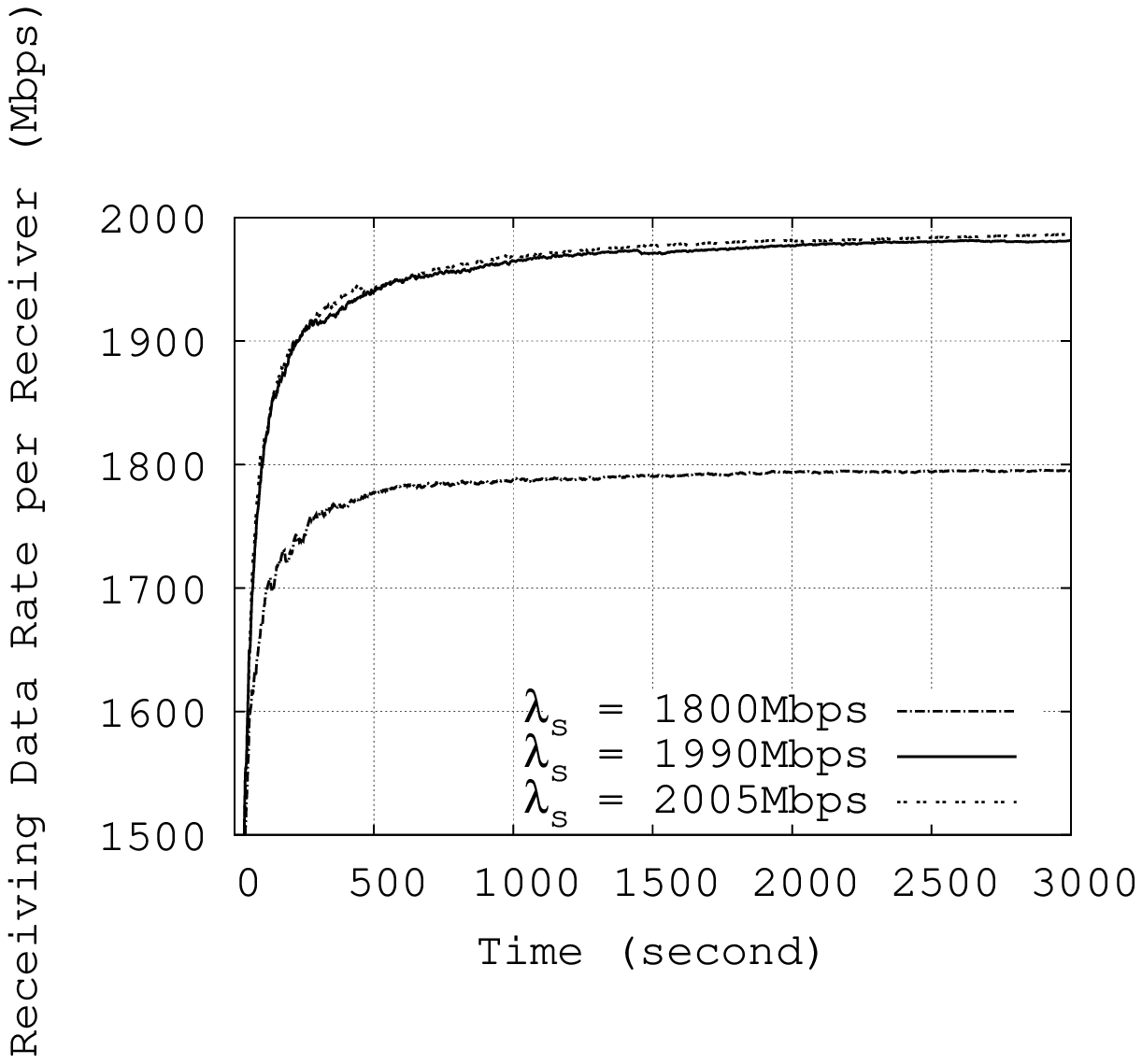} \\ (b)
\end{center}
\end{minipage}
\end{center}
\caption{Convergence of the two algorithms;
(a) the first algorithm;
(b) the second algorithm.}
\label{fig:receiving-rate}
\end{figure}

Fig. \ref{fig:StabilitySecondAlgorithm} shows that both the network virtual queues and the real queues
remain stable if $\lambda_s$ is below the maximum
achievable session rate.
When $\lambda_s = 1990$ Mbps, the aggregate real queue size of all the network queues
is under $25$ GB, which is about
the amount of data flowing through the network in a $5$-second interval.
Under the same arrival rate, the average queue size per link is $183.8$ MB and the largest real queue size at a link is $1.8$ GB, which is $2.88$ times of the bandwidth-delay product of a critical link. For $\lambda_s = 1800$ Mbps, the aggregate queue size is under $15$ GB, the average queue size per link is under $110.3$ MB, and the maximum queue size at a link is
under $1$ GB. Hence, the queue sizes can be made much smaller with a slightly reduced arrival rate.
Compared with the first algorithm, a larger buffer is required at each link.
When $\lambda_s$ exceeds the maximum achievable session rate, both the aggregate virtual queue size and the aggregate real queue size grow indefinitely; the network is unstable.
After reaching the steady state, the real queue process exhibits more oscillation than that in the first algorithm. This is in part because the randomly selected tree on each time slot is not necessarily the min-cost tree. The aggregate queue size oscillates within a $10$-GB range and the average oscillation is $73.5$ MB at each link. Fig. \ref{fig:receiving-rate} (b) shows that the average receiving rate converges reasonably fast to the arrival rate when the arrival rate is below the maximum achievable session rate.


\subsection{Results for the Larger Network}

\begin{figure}[t]
\begin{center}
\begin{minipage}{3.2in}
\begin{center}
\includegraphics[width=3.2in]{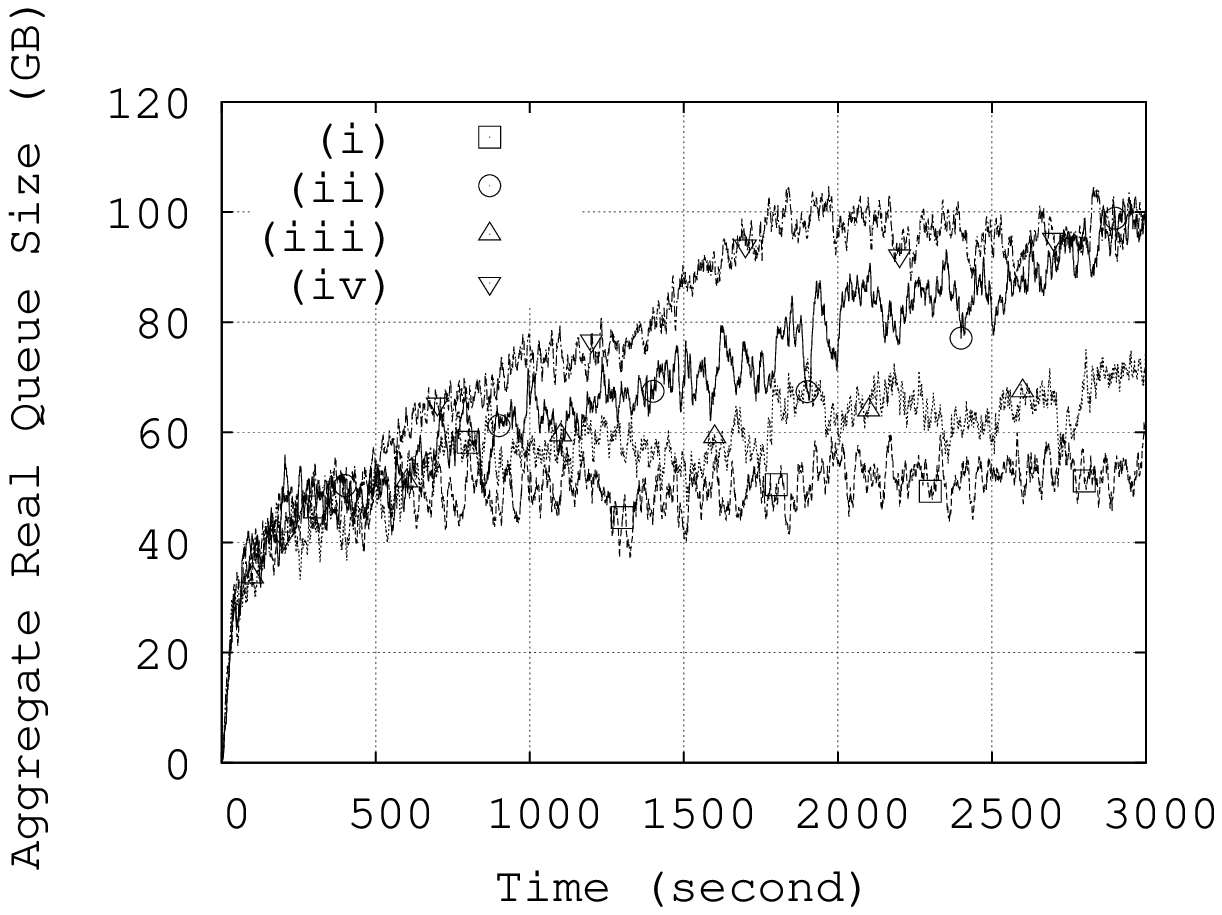} \\ (a)
\end{center}
\end{minipage}
\begin{minipage}{3.2in}
\begin{center}
\includegraphics[width=3.2in]{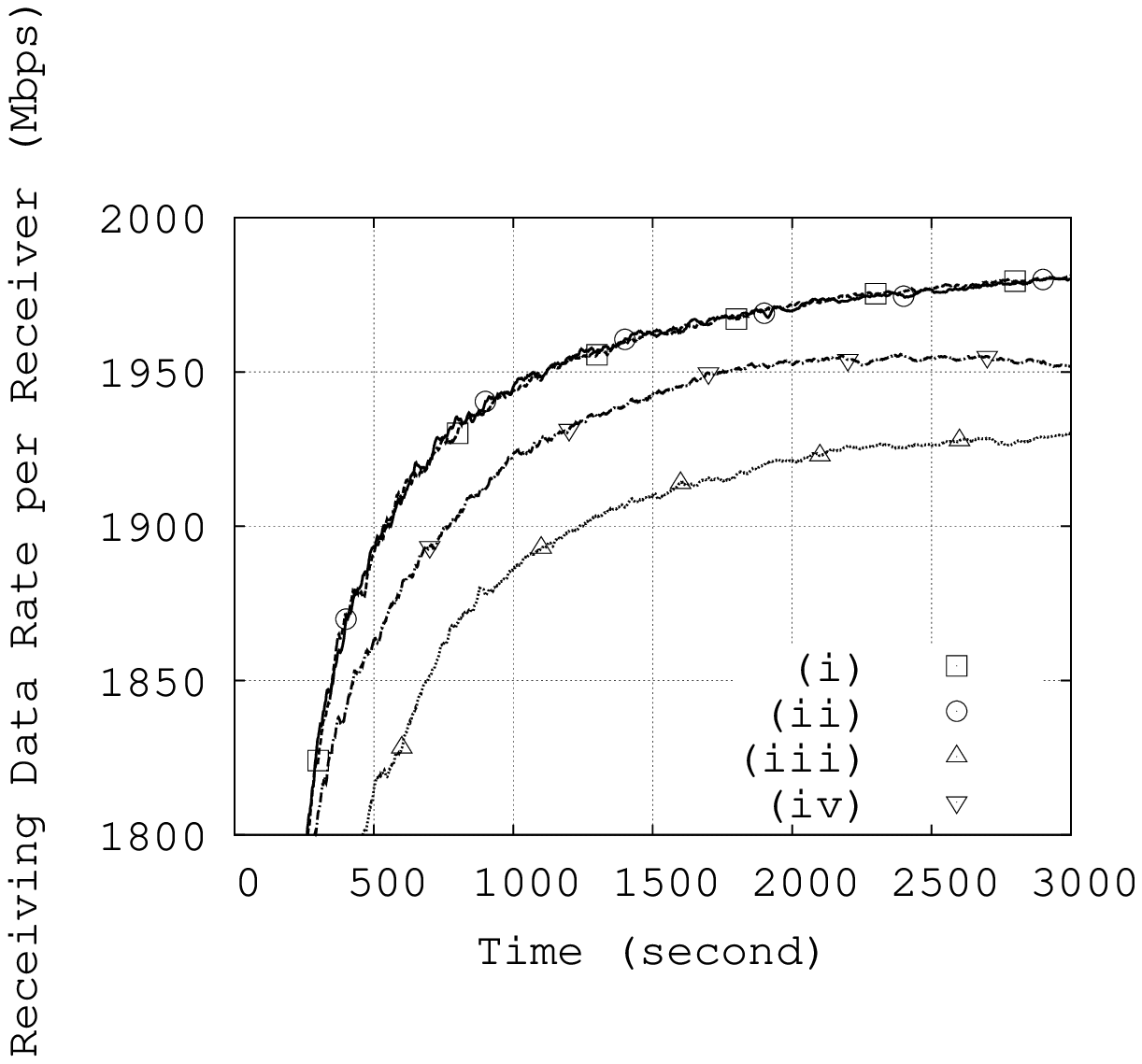} \\ (b)
\end{center}
\end{minipage}
\end{center}
\caption{The experiments on a network with 295 nodes and 1086 links;
(a) aggregate real queue sizes; (b) average receiving rate per receiver. Legends:
(i) the first algorithm with $\lambda_s = 2$ Gbps,
(ii) the first algorithm with $\lambda_s = 2.1$ Gbps,
(iii) the second algorithm with $\lambda_s = 2$ Gbps,
(iv) the second algorithm with $\lambda_s = 2.1$ Gbps.
}
\label{fig:larger_network}
\end{figure}

We have conducted more experiments on the larger ISP network with $295$ nodes and $1086$ links. We assign $5$ Gbps link capacity to each of the critical links and $1$ Gbps to each of the other links,
and we set up one multicast session with $39$ receivers.
The experiments suggest that the maximum session rate is around $2$ Gbps, although we do not know the exact value.
Fig. \ref{fig:larger_network} (a) shows that the network is stable when
the source rate $\lambda_s$ is $2$ Gbps and the real queues build up indefinitely
when the source rate is $2.1$ Gbps. For the case of $\lambda_s = 2$ Gbps, the aggregate queue size is under $60$ GB or $80$ GB for the first or second algorithms, respectively; the average queue size per link is under $55.2$ MB or $73.7$ MB, respectively. The oscillation of the aggregate queue size in the steady state is under $15$ GB for both algorithms; after divided by the number of links, the average is under $13.8$ MB per link.
The largest queue size observed at a link is $3$ GB for the first algorithm and $21$ GB for the second algorithm. Evidently, tree selection in the first algorithm is better at avoiding congested links. In the second algorithm, a link coming out of the source has the largest queue size for much of the simulation duration. Better tree selection should help the second algorithm to reduce the largest queue size. Overall, these are promising results for buffer requirements, particularly considering the fact there are more receivers in this set of experiments. Fig. \ref{fig:larger_network} (b) shows that, in each stable case where $\lambda_s=2$ Gbps, the average receiving
rate per receiver ramps up to $1.8$ Gbps in under $300$ time slots, which is $90\%$ of the arrival rate. Again, this receiving rate is a time average as well. The convergence speed of the instantaneous receiving rate can be inferred from Fig. \ref{fig:larger_network} (a).
There, we see that the queues become steady in less than $500$ or $1000$ time slots for the first and second algorithm, respectively. We can deduce that convergence to greater than $99\%$ of the final value has been achieved within those time slots. We conclude that the algorithms converge fairly fast.

\subsection{Control Overhead}

We will briefly describe a simple design of the signaling/control protocol and show that the control overhead is small.
On each time slot, a source sends one signaling packet to each node on the multicast tree selected by the algorithm. The signaling packet directed toward a node contains a list of link IDs (32 bits each), which designate the node's outgoing links on the multicast tree. The packet also contains a 32-bit virtual source rate and a 32-bit multicast session ID (which can use the source ID). The second type of control packets has the function of carrying the costs of the links (i.e., links' virtual queue sizes) back to the sources of the multicast sessions. On each time slot, a node sends one feedback packet to each source. The packet contains the IDs and costs (32 bits each) of the node's outgoing links and the 32-bit node ID. Every control packet has an additional 20-byte header containing miscellaneous information. The size of each control packet is well under $1000$ bytes in typical cases.

For a network with $n$ nodes and $m$ links, the total size of all the signaling packets from a source is at most $(20 \times 8 + 32 + 32) n + 32 m$ bits on each time slot. Consider a fairly aggressive example where the network has $300$ nodes and $3000$ links and the time slot size is $0.5$ seconds. On each time slot, the total size of the forward signaling packets is $163,200$ bits, which means the control traffic rate from the source is $326.4$ Kbps. If the source data rate is $100$ Mbps, the forward signaling overhead is only $0.3264 \%$. Furthermore, the control traffic rate is independent of the source data rate. As technology progresses to $1$ Gbps or $10$ Gbps source rates, the overhead becomes $0.03264\%$ or $0.003264 \%$, respectively. It is also sensible to define the signaling overhead with respect the {\em total} data traffic rate for the multicast session. For a multicast session with $r$ receivers, the total data traffic rate entering the receivers is equal to $r$ times of the source rate. If a multicast session has $100$ receivers and the source rate is $100$ Mbps, the total data rate is $10$ Gbps. The signaling traffic is $0.003624\%$ of $10$ Gbps.

The second type of control packets has the function of carrying the costs of the links (i.e., links' virtual queue sizes) back to the sources of the multicast sessions. On each time slot, each node sends one feedback packet to every source. The packet contains the IDs and costs of the node's outgoing links, the node ID and a 20-byte header. Then, the total size of all the feedback packets received by a source on each time slot is $(20 \times 8 + 32)n + (32+32)m$ bits. For a network with $300$ nodes and $3000$ links, the total size is $249,600$ bits, which corresponds to a traffic rate of $499.2$ Kbps when the time slot size is 0.5 seconds. For $100$ Mbps, $1$ Gbps or $10$ Gbps source rates, the overhead is $0.4992\%$, $0.04992\%$ or $0.004992\%$, respectively, when compared with the source rate; it is $0.004992\%$, $0.0004992\%$ or $0.00004992\%$, respectively, when compared with the total data rate for a session with $100$ receivers.


\section{Related Work}
\label{sec:related}

Research on similar stability questions has been very active, but
generally, in the context of unicast (e.g. \cite{TE92,
Tass95, ES05, Neely-IT06, GNT06, LS06A, WSP07,
JLS08}), possibly with multiple paths per connection.
The presence
of multicast puts our problem in a class of its own in that many
earlier stable control algorithms, such as the maximum
backpressure-based algorithm \cite{Tass95}, and techniques
for stability analysis are not directly applicable. The main
reason is that, unlike unicast, the flow conservation condition no
longer holds under multicast.

\eat{
\textcolor{red}{
There are a few works about the stability analysis of
multicast/broadcast \cite{MassoulieTGR07, BonaldMMPT08,
KumarYR07, MundingerWW08, SanghaviHM07, DanFC07},
where most of them assume an access constrained network
with the only exception of \cite{MassoulieTGR07}.
In \cite{BonaldMMPT08, KumarYR07,
MundingerWW08, SanghaviHM07},
various Bittorrent-like algorithms are proposed and
are proved to achieve the optimal performance in
terms of distribution rate and/or delay, where
the bottleneck is at the upload links with
heterogeneous capacity.
In \cite{MassoulieTGR07}, Massoulie et al. presented a simple
local-control algorithm for broadcasting
in arbitrary networks, which provably achieves
the optimal broadcast rate. The algorithm only
allows broadcast for a single source and requires
all nodes in the network to participate in receiving
the complete copy of data.
In \cite{DanFC07}, the stability performance of multiple-tree-based
peer-to-peer live streaming is analyzed, where
the stability is defined as the availability of
data in presence of node dynamics.
}
}

There are several papers on the stability analysis of
multicast/broadcast \cite{MassoulieTGR07, BonaldMMPT08,
KumarYR07, SanghaviHM07, DanFC07}. Except \cite{MassoulieTGR07}, most of these
assume an access constrained network.
In \cite{BonaldMMPT08, KumarYR07, SanghaviHM07},
various Bittorrent-like algorithms are proposed and
are proved to achieve the optimal performance in
terms of the distribution rate and/or delay, where
the bandwidth bottleneck is at the upload links.
In \cite{MassoulieTGR07}, Massouli\'{e} {\em et. al.} present a simple
local-control algorithm for broadcast
in a general network, which provably achieves
the optimal broadcast rate. The algorithm only
allows broadcast from a single source and requires
all nodes in the network to receive a
complete copy of the data.
In \cite{DanFC07}, the stability of multiple-tree-based
peer-to-peer live streaming is analyzed, where
stability is defined as the availability of
data in the presence of node dynamics.

Another salient aspect of the universal swarming problem is most
related to the problem of link scheduling in wireless networks
subject to link interference constraints, which has attracted much
attention recently \cite{TE92, Tassiulas98, JX06,
LS06A, CLCD06, Neely-IT06, DimakisW06,
SSM07, GLS07, JLS08}. In \cite{TE92}, Tassiulas {\em et.
al.} showed that the maximum-weight link schedule achieves (i.e.,
stabilizes) the interior of the throughput
region, where
the weights are the queue size differences, or the backpressure.
However, finding such a schedule is in general an NP-hard problem.
The universal swarming problem usually involves an NP-hard
subproblem in order to achieve the entire throughput region, which
is to find a minimum-cost Steiner tree. This similarity makes many
of the concerns and investigative approaches in the wireless link
scheduling problem relevant to the universal swarming problem. In
\cite{LS06A, LS06B}, Lin {\em et. al.} showed that approximation
algorithms for the maximum-weight scheduling problem can be used
to stabilize a portion of the throughput region. Some researchers
considered maximal scheduling algorithms and studied what their
stability regions are \cite{DimakisW06, WSP07, SSM07, GLS07, JLS08}. Other authors proposed randomized scheduling algorithms that achieve the entire throughput region \cite{Tassiulas98} \cite{Ra09}.

\section{Conclusions and Discussions}
\label{sec:conclusion}

In universal swarming, packets are distributed to the receivers
along multiple multicast trees. The paper focuses on analyzing the
stability of the algorithms for sending dynamically arriving
packets onto the trees. To achieve the throughput region, we
encounter a min-cost Steiner tree problem, which is NP-hard.
Multi-hop traffic is another difficulty for finding stable
universal swarming algorithms. We propose a $\gamma$-approximation
min-cost tree scheduling algorithm with network signaling and
source regulators. It guarantees network stability in a reduced
throughput region, where the reduction ratio is no more than the
approximation ratio of the algorithm for the min-cost tree
problem. We further develop a randomized tree-scheduling algorithm
with network signaling. It achieves the throughput region and
stabilizes the virtual queues. Moreover, the real queue processes are rate stable and the average traffic
intensity at each link is strictly less than one. However, whether
or not the real queues are stable remains an open question.

In the worst case, even finding an approximate min-cost tree can be very time consuming. However, the algorithms and results in this paper can still be practically relevant. First, our algorithms do not require to find the true min-cost tree. In practice, there is a time budget (e.g., the time slot size) in the tree computation step. The tree that is found to have the least cost during the budgeted time will be used by our algorithms. Second, there are several possibilities that, in practice, the tree-computation time may not be prohibitively long. (i) In small networks (less than 100 nodes and links), finding the min-cost tree can be quite fast. (ii) For the intended application, the network topology is a fixed one, not an arbitrary one. One may be able to find specialized, fast algorithms for that particular topology. (iii) For a fixed topology, one may be able to find a heuristic algorithm that has near optimal performance most of the time. (iv) In many practical networks, such as the ones used in our simulation, it is sometimes quite clear that a small number of links are the critical ones and they almost always experience large queue sizes. It is not necessarily hard to find a min-cost (or nearly min-cost) tree, if only these critical links need to be inspected. Finally, the tree computation time can be reduced drastically at the expense of a small reduction in the throughput. If the time budget for tree computation is chosen to be very short, e.g., close to the round-trip propagation time, we may restrict each multicast session to use a small number of precomputed trees (e.g., 4 -- 100 trees); finding the min-cost tree among them is trivial. Experiences have shown that the loss in throughput due to the restriction on the trees is likely to be small (see \cite{RRX08} for the unicast case). The performance guarantees found in the paper still apply if we modify the optimization problem by adding the constraints about the allowed trees.

We now briefly discuss how to adapt the algorithms to the case of small buffer sizes. This is possible because, fundamentally, the algorithms use the virtual queue sizes for rate computation and they will find the optimal rates regardless of the actual buffer sizes. The first idea is again to use a small number of pre-computed trees for each multicast session, say $L$ trees. Since finding the least-cost tree among the $L$ trees is extremely easy, we can reduce the time slot size to near the round-trip propagation time, and hence, reduce the bandwidth-delay product.
Next, we wish to reduce temporary overload in the steady state due to tree-hopping. The idea is to use the $L$ multicast trees {\em simultaneously} with the correct tree rates.
In the modified algorithm, every source computes the time average of the virtual source rates (for releasing virtual packets to the network) that each of its $L$ trees sees; this yields the time averages of the virtual tree rates and these time-average rates converge to the optimal tree rates regardless of the buffer size\footnote{Here, the optimal rate-allocation problem is a modified one. Compared with the original optimization problem, the only modification is that it assumes $L$ fixed trees per multicast session.}. The computed average rates are used as targets for what the real tree rates (and hence, the real source rate) should be. In a static network environment (i.e., one with fixed topology, constant link bandwidth and long-lasting multicast sessions), the real tree rates can be set directly at the time-average virtual tree rates when the latter have stabilized\footnote{The initial waiting time for the convergence of the time-average virtual rates is not a concern under the static assumption.}. Then, the only remaining possible cause for queue buildup is the randomness in the arrival process. A very small buffer size is sufficient to absorb such traffic fluctuation and prevent packet losses (e.g., $10$-$100$ packets).
Finally, we can further enhance the algorithms to reduce the transient buildup of the queues so that the algorithms can cope with macro-level dynamics, by which we mean changes in the network topology and link bandwidth or the arrivals/departures of multicast sessions. The idea is to have another level of adaptation by the real tree rates. For instance, a real tree rate can rapidly increase to $80\%$ of the time average of the corresponding virtual tree rate; after that, it increases gradually until packet losses occur, at which point, it drops to half of the time-average virtual tree rate. This way, the real rates will not overshoot too much. Note that the adaptation of the real rates and the computation of the time averages are done concurrently; there is no waiting for the convergence of the time-average rates.

\appendices

\section{Proofs}
\label{sec:appendix}

\subsection{Proof of Theorem \ref{theorem:Q_bound_regulator}}
Recall that every packet at any link has a hop count from the source, which
is the hop count on its tree path to the link. If a packet has a hop count
$h$ when it arrives at a link, we say it belongs to the $h^{th}$-hop traffic.
Let $Q_e(k,h)$ denote the queue backlog at link $e$ at time $k$
contributed by all first-hop through $h^{th}$-hop traffic. We assume $Q_e(k, 0) = 0$ for ease of presentation.


For each $e \in E$,
let $\bar{h}_e \leq |V| - 1$ be the largest hop count of any packet in $e$.
Let $\bar{h} = \max_{e \in E} \bar{h}_e$.
We claim that for all $h = 1, \cdots, \bar{h}$,
\be \label{ineq:hop_Q_bound}
Q_e(k, h) \leq M_h, \ \ \ \forall k, \ \forall e \in E,
\ee
where the constants satisfy
$M_1 \leq M_2 \leq \dots \leq M_{\bar{h}} < \infty$.
We prove this claim by induction.

\noindent {\bf Base}: When $h = 1$, note that
$Q_e(k, 1) \leq q_e(k)$ and $q_e(k) \leq M_e$ by Corollary \ref{corollary:q_bound_regulator}.
Let $M_1 = \max_{e \in E} M_e$. Then, (\ref{ineq:hop_Q_bound}) holds for $h = 1$ and for all $k$.

Assume (\ref{ineq:hop_Q_bound}) holds for
$1, \cdots, h-1$ and for all $k$.

%

\noindent {\bf Induction on $h$}:
Let $x_e(k, h)$ denote the number of packets arriving at link
$e$ on time slot $k$ that belong to the first-hop through $h^{th}$-hop traffic.
Then, during any interval of time, $k_0$ through $k$,
the total number of arrivals is
no more than the sum of the number of packets released by the sources
during this interval
that travel through link $e$
and all the backlogged, first-through-$(h-1)^{th}$-hop packets in the network
at time $k_0$, i.e.,
\begin{align}
 \sum_{u = k_0}^k x_e(u, h)
\leq & \sum_{u = k_0}^k \sum_{s \in S}
    D_s(u) I(e, t_s(u)) + \sum_{e' \in E} Q_{e'} (k_0, h - 1)
    \nonumber \\
\leq & \sum_{u = k_0}^k \sum_{s \in S}
    D_s(u) I(e, t_s(u)) + |E| \cdot M_{h - 1}. \label{ineq:x_interval}
\end{align}
(\ref{ineq:x_interval}) holds due to the induction hypothesis.

Assume all the queues are empty at time 0. For all $k$,
\begin{align}
\label{eq:Loynes}
Q_e(k, h)
= \ & \max_{0 \leq k_0 \leq k}
    \{ \sum_{u = k_0}^{k} x_e(u, h) - c_e (k - k_0 + 1) \} \\
\leq \ & \max_{0 \leq k_0 \leq k}
    \{ \sum_{u = k_0}^k \sum_{s \in S}
    D_s(u) I(e, t_s(u))
    + |E| \cdot M_{h - 1}
    - c_e (k - k_0 + 1) \} \label{eq:loy2} \\
\leq \ & M_e + |E| \cdot M_{h - 1}. \label{eq:loy3}
\end{align}
(\ref{eq:Loynes}) is by
an elementary fact about the queue size. It is derived by expanding the one-step queue update backward in time. It says that the queue size is
the largest difference between the total arrivals and the accumulated link capacity on any sub-interval ending at the current time $k$.
(\ref {eq:loy2}) is by using (\ref{ineq:x_interval})
in (\ref{eq:Loynes}).
(\ref{eq:loy3}) is by Lemma \ref{lemma:link_arrival_bound_regulator}.
We can define $M_{h} = M_e + |E| \cdot M_{h-1}$.

Finally,
note that the overall queue backlog $Q_e(k)$ at link $e$ is equal to
$Q_e(k, \bar{h}_e)$, where $\bar{h}_e \leq \bar{h}$.
Hence, $Q_e(k) \leq M_{\bar{h}_e} \leq M_{\bar{h}}$ for all $k$.

\subsection{Proof of Theorem \ref{theorem:q_ergodicity_regularity}}
The proof follows from Theorem 8.0.3 in \cite{Meyn08}. Denote by
$\mathcal{X}$ the state space of the Markov process $\{ x(k) \} =
\{q(k), t(k)\}$. Define a finite subset of the state space
$\hat{\mathcal{X}} = \{x \in \mathcal{X}: \sum_{e \in E} q_e \leq
\frac{M}{\epsilon} \}$, for $M$ and $\epsilon$ as specified in
Lemma \ref{lemma:lyapunov_drift_random}. Note that by assumption
AS \ref{as:formarkov}, the process $\{x(k)\}$ is an
$x^*$-irreducible Markov chain with $x^* = (q^*, t^*)$, where $q^*
= \vec{0}$ and $t^*$ is a fixed set of  minimum-cost trees under
the link cost vector $q^* = \vec{0}$;\footnote{There might be
multiple sets of the minimum-cost trees. Let $t^*$ be an arbitrary
one of them. Since the tie is broken uniformly at random, there is
a non-zero probability that we choose the set $t^*$.} and it is
also aperiodic on the countable state space $\mathcal{X}$.
Assumption AS \ref{as:formarkov} says that on every time slot,
there is some nonzero probability that no new arrivals enter the
sources. The system will empty after a finite number of such
successive ``no arrival'' slots, an event that has a positive
probability. This implies the process $\{x(k)\}$ is an
$x^*$-irreducible Markov chain (i.e., $\sum_{k = 0}^{\infty}
P^k(x, x^*) > 0, x \in \hat{\mathcal{X}}$). Aperiodicity follows
from $P(x^*, x^*) = P\{A(k) = 0 \} \cdot P\{t^* \mbox{ is
chosen}\} > 0$.

By Lemma \ref{lemma:lyapunov_drift_random}, the chain $\{x(k)\}$
satisfies the Foster-Lyapunov drift conditions \cite{Meyn08}. Now
all the conditions of Theorem 8.0.3 in \cite{Meyn08} are met.
Hence, Theorem \ref{theorem:q_ergodicity_regularity} holds.

\subsection{Proof of Theorem \ref{theorem:cumulative_arrival}}
According to (\ref{eq:q_update}),
\be
q_e(k + 1) \geq q_e(k) + \sum_{s \in S} A_s(k) I(e, t_s(k)) - (c_e - \epsilon_2),
\forall e \in E. \nonumber
\ee
Summing the above inequality from time slots $0$ to $k$, we have
\begin{align}
\label{ineqs:cumulative_arrival_proof}
 \sum_{u = 0}^k \sum_{s \in S} A_s(u) I(e, t_s(u))
\leq \ & (c_e - \epsilon_2)(k + 1) + q_e(k + 1) - q_e(0) \nonumber \\
\leq \ & (c_e - \epsilon_2)(k + 1) + q_e(k + 1).
\end{align}
Dividing both sides of (\ref{ineqs:cumulative_arrival_proof}) by
$k+1$ yields
\begin{align}
 \frac{\sum_{u = 0}^k \sum_{s \in S} A_s(u) I(e, t_s(u))}{k+1}
\leq \ & c_e - \epsilon_2 + \frac{q_e(k + 1)}{k+1}         \nonumber \\
= \ & c_e - \epsilon_2 +
\frac{\sum_{u = 0}^{k+1} q_e(u) - \sum_{u = 0}^k q_e(u)}{k+1}   \nonumber \\
= \ & c_e - \epsilon_2 +
\frac{\sum_{u = 0}^{k+1} q_e(u)}{k+2} \cdot \frac{k+2}{k+1}
- \frac{\sum_{u = 0}^k q_e(u)}{k+1}.   \nonumber
\end{align}
Taking the limit on both sides of the above inequality as $k$ goes
to infinity, we get,
\begin{align}
 \limsup_{k \rightarrow \infty}  \frac{\sum_{u = 0}^k \sum_{s \in S}
    A_s(u) I(e, t_s(u))}{k+1}
\leq \ & c_e - \epsilon_2 +
\lim_{k \rightarrow \infty} \frac{\sum_{u = 0}^{k+1} q_e(u)}{k+2}
\cdot \lim_{k \rightarrow \infty} \frac{k+2}{k+1}  \nonumber \\
\ & \quad - \lim_{k \rightarrow \infty} \frac{\sum_{u = 0}^k q_e(u)}{k+1} \nonumber \\
= \ & c_e - \epsilon_2. \nonumber
\end{align}
The last equality holds because $\lim_{k \rightarrow \infty}
\frac{\sum_{u = 0}^{k+1} q_e(u)}{k+2} = \lim_{k \rightarrow
\infty} \frac{\sum_{u = 0}^k q_e(u)}{k+1} = E[\hat{q}_e]$ by Theorem
\ref{theorem:q_ergodicity_regularity}. Hence,
(\ref{ineq:cumulative_arrival_strong_law}) holds.

Now taking the expectation on the both sides of
(\ref{ineqs:cumulative_arrival_proof}) yields
\begin{align}
E[\sum_{u = 0}^k \sum_{s \in S} A_s(u) I(e, t_s(u))]
& \leq (c_e - \epsilon_2)(k + 1) + E[q_e(k + 1)] \nonumber \\
& \leq (c_e - \epsilon_2)(k + 1) + M_e,  \nonumber
\end{align}
where $E[q_e(k + 1)] \leq M_e < \infty$.
Because Theorem \ref{theorem:q_ergodicity_regularity} says,
$\lim_{k \rightarrow \infty} E [q_e(k)] \rightarrow E[\hat{q}_e]$
and $E[\hat{q}_e] < \infty$, such $M_e$ exists.
Dividing both sides of the above inequality by $k+1$ and
taking the limit, we have,
\begin{align}
& \limsup_{k \rightarrow \infty}
    E[\frac{1}{k+1}\sum_{u = 0}^k \sum_{s \in S} A_s(u) I(e, t_s(u))]
\leq c_e - \epsilon_2 + \lim_{k \rightarrow \infty} \frac{M_e}{k+1}
= c_e - \epsilon_2. \nonumber
\end{align}


\subsection{Proof of Corollary \ref{corollary:Q_rate_stable}}
For any time slot $k$, let $\{a_e(k)\}$ denote the real
packet arrival process at link $e$.
The real queue dynamic equation can be written as
\be
Q_e(k + 1) = \max [Q_e(k) - c_e, 0] + a_e(k).
\ee
Note that, for any time slot $k$, the number of cumulative arrivals at the real queue is no
more than the number of cumulative arrivals at the virtual queue,
i.e.,
\be
\nonumber
\sum_{u = 0}^k a_e(u)
\leq
\sum_{u = 0}^k \sum_{s \in S} A_s(u)I(e, t_s(u)).
\ee
Then, by Theorem \ref{theorem:cumulative_arrival},
\begin{align}
\limsup_{k \rightarrow \infty} \frac{\sum_{u = 0}^k a_e(u) }{k}
\leq
\limsup_{k \rightarrow \infty}
\frac{\sum_{u = 0}^k \sum_{s \in S} A_s(u)I(e, t_s(u))}{k}
\leq c_e - \epsilon_2 \leq c_e.
\nonumber
\end{align}
According to the Rate Stability Theorem in \cite{Neely2010} (Theorem 2.4),
$Q_e(k)$ is rate stable, i.e.,
\be
\lim_{k \rightarrow \infty} \frac{Q_e(k)}{k} = 0 \ \
\mbox {with probability } 1.
\nonumber
\ee

\bibliographystyle{IEEEtran}
\bibliography{Stability}

\begin{IEEEbiography}
{Xiaoying Zheng}
received the bachelor's and master's degrees in computer science and engineering
from Zhejiang University, P.R. China, in 2000 and 2003, respectively, and the
PhD degree in computer engineering from the University of Florida, Gainesville, in 2008.
She is an assistantXiaoying Zheng received the bachelor’s and master’s degrees in computer science and engineering from Zhejiang University, P.R. China, in 2000 and 2003, respectively, and the PhD degree in computer engineering from the University of Florida, Gainesville in 2008. She was an assistant professor in Shanghai Research Center of Wireless Communications, from March 2009 to October 2011. She then joined Shanghai Advanced Research Institute, Chinese Academy of Sciences in 2012 as an associate professor. Her research interests include applications of optimization theory in networks, distributed systems, performance evaluation of network protocols and algorithms, peer-to-peer overlay networks, content distribution, and congestion control.  professor at Shanghai Research Center for Wireless Communications
and Key Lab of Wireless Sensor Network and Communications,
Chinese Academy of Sciences (CAS), China.
Her research interests include applications of optimization theory
in networks, performance evaluation of network protocols and
algorithms, peer-to-peer overlay networks, wireless networks,
content distribution and congestion control.
\end{IEEEbiography}

\begin{IEEEbiography}
{Chunglae Cho}
received the B.S. and M.S. degrees in computer science from Pusan National University, Korea, in 1994 and 1996, respectively. He worked as a research staff member at Electronics and Telecommunications Research Institute, Korea, between 2000 and 2005. He is currently working toward a Ph.D. degree at the Computer and Information Science and Engineering department at the University of Florida, Gainesville, FL. His research interests are in computer networks, including resource allocation and optimization on peer-to-peer networks and wireless networks.
\end{IEEEbiography}

\begin{IEEEbiography}
{Ye Xia} is an associate professor at the Computer and Information
Science and Engineering department at the University of Florida,
starting in August 2003. He has a PhD degree from the University
of California, Berkeley, in 2003, an MS degree in 1995 from
Columbia University, and a BA degree in 1993 from Harvard
University, all in Electrical Engineering. Between June 1994 and
August 1996, he was a member of the technical staff at Bell
Laboratories, Lucent Technologies in New Jersey. His main research
area is computer networking, including performance evaluation of
network protocols and algorithms, resource allocation, wireless
network scheduling, network optimization, and load balancing on
peer-to-peer networks. He also works on cache organization and
performance evaluation for chip multiprocessors. He is interested
in applying probabilistic models to the study of computer systems.
\end{IEEEbiography}

\end{document}